\documentclass[final,1p,times]{elsarticle}

\usepackage{amssymb}

\usepackage{subfig}
\usepackage{comment}
\usepackage{tikz}
\usepackage{nicefrac}
\usepackage{amsmath}
\usetikzlibrary{shapes.geometric} 
\usetikzlibrary[shapes.geometric] 
\usetikzlibrary{shapes}
\usetikzlibrary{patterns}
\usetikzlibrary{plotmarks}

\journal{Nuclear Instruments and Methods A}

\begin{document}

\begin{frontmatter}



\title{Cryogenic magnetic coil and superconducting magnetic shield for neutron electric dipole moment searches}


\author[CIT]{S. Slutsky}
\author[CIT]{C. M. Swank}
\author[CIT]{A. Biswas}
\author[CIT]{R. Carr}
\author[CIT,X]{J. Escribano}
\author[CIT]{B. W. Filippone}
\author[CIT,US]{W. C. Griffith}
\author[CIT,LLNL]{M. Mendenhall}
\author[UK]{N. Nouri}
\author[CIT]{C. Osthelder}
\author[CIT,Vertex]{A. P\'{e}rez Galv\'{a}n}
\author[CIT,Triumf]{R. Picker}
\author[UK]{B. Plaster}

\address[CIT]{Department of Physics, Math and Astronomy, California Institute of Technology, Pasadena, CA 91125 USA}
\address[X]{Xamarin, San Francisco, CA 94111 USA}
\address[US]{Department of Physics and Astronomy, University of Sussex, Brighton, BN1 9RH United Kingdom}
\address[LLNL]{Nuclear and Chemical Sciences Division, Lawrence Livermore National Laboratory, Livermore, CA 94550, USA}
\address[UK]{Department of Physics and Astronomy, University of Kentucky, Lexington, KY 40506 USA}
\address[Vertex]{Vertex Pharmaceuticals, 11010 Torreyana Rd., San Diego, CA 92121 USA}
\address[Triumf]{TRIUMF, Vancouver, BC V6T 2A3, Canada}

\begin{abstract}
A magnetic coil operated at cryogenic temperatures is used to produce spatial, relative field gradients below 6 ppm/cm, stable for several hours. The apparatus is a prototype of the magnetic components for a neutron electric dipole moment (nEDM) search, which will take place at the Spallation Neutron Source (SNS) at Oak Ridge National Laboratory using ultra-cold neutrons (UCN). That search requires a uniform magnetic field to mitigate systematic effects and obtain long polarization lifetimes for neutron spin precession measurements. This paper details upgrades to a previously described apparatus \cite{Adrian}, particularly the introduction of super-conducting magnetic shielding and the associated cryogenic apparatus. The magnetic gradients observed are sufficiently low for the nEDM search at SNS.    
\end{abstract}




\end{frontmatter}


\section{Introduction}
\label{sec:Intro}

The existence of a permanent electric-dipole-moment (EDM) on a subatomic scale would violate both parity (P) and time (T) symmetries, and would be a signature of physics beyond the Standard Model \cite{RamseyEDMReview}. Additionally, with CPT symmetry, such an EDM would also violate the combined charge and parity (CP) symmetry. The amount of CP violation currently observed in meson decay cannot explain the baryonic matter/anti-matter asymmetry in the observed universe, within the Sakharov criteria \cite{Sakharov}. Therefore, a larger source of CP violation is anticipated. 
With the expected precision of the next generation of competitive EDM searches, a positive or null measurement will have broad theoretical consequences \cite{ChuppRM, JungPich, IORM}.

Of the species available to probe EDMs, the neutron has several advantages. Relative to atoms or charged particles, the neutron can be considered simple to understand and manipulate. It has no electrons to shield or enhance the effect of an EDM, mitigating errors that can arise from theoretical predictions. Its trajectory is not affected by uniform electric fields, and it can be trapped in a material bottle 
and observed for periods only limited by its intrinsic lifetime. Currently, the most precise nEDM measurement was made using ultracold neutrons (UCN) by the Sussex/RAL/ILL nEDM experiment, setting a limit of $3.0\times10^{-26}$ e-cm \cite{ILL}. Among the next generation of EDM experiments, the nEDM search at the SNS \cite{ClaytonLeptonMoments2014}, based on the concepts discussed in \cite{GolubLamoreauxPhysicsReports}, is possibly the most ambitious of all, with a design sensitivity of $d_n\lesssim3\times10^{-28}$ e-cm. 


While the Sussex/RAL/ILL experiment was statistically limited, the largest systematic uncertainty was due to the so-called ``geometric phase'' \cite{GeoPhase}. When a particle's spin precession frequency is measured in the presence of a magnetic field with a spatial gradient, there is a frequency shift proportional to both the linear magnetic field gradients and to the applied electric field; this can mimic the expected EDM signal. It is therefore critical to demonstrate that field gradients are under control and understand what materials constitute potential sources of magnetic field before designing a full-scale experiment. To that end, we constructed a prototype replicating the magnetic coils and shielding of the SNS experiment at half-scale in each linear dimension, and we evaluated the magnetic fields inside. 

\section{Experimental apparatus}
\label{sec:Apparatus}

\subsection{Summary of the nEDM apparatus at the SNS}


In the SNS nEDM experiment, neutrons will be generated at the mercury spallation source and moderated to low temperatures, $\sim$20-30K. The cold neutrons will be spin-polarized with a supermirror polarizer and then guided into the cryogenic apparatus, where they will illuminate two cells filled with isotopically pure $^4$He. The two cells will be held at a temperature of $\sim$450 mK by a large dilution refrigerator. Neutrons with a wavelength of 8.9 Angstroms can interact with the superfluid via phonon emission and down-scatter to an energy of $\sim$0-200 neV \cite{SuperThermalUCN1, SuperThermalUCN2}. These neutrons will be moving so slowly that they become trapped between the walls of the cell, which are coated with deuterated polystyrene. The neutrons that become trapped are considered UCN. 

Trapped UCN will be subjected to a magnetic holding field, $B_{0} \sim 3\mu$T, to maintain polarization. A strong electric field, $E$, will be applied to probe the EDM, the direction of which can be reversed to control for systematics. A $\pi$/2 pulse will rotate the UCN spins perpendicular to the holding field. The UCN spins will then precess according to their Larmor frequency, $\omega_{n}$: 
\begin{align}
\omega_{n} =  -2(\mu_{n}B_{0} \pm d_{n}E)/\hbar \label{eq:larmor}
\end{align} 
where $\mu_{n}$ and $d_{n}$ are the neutron magnetic and electric dipole moments, respectively, and $\hbar$ is Planck's constant. The sign of the EDM term depends on the direction of the applied field. Thus, a neutron spin-precession frequency shift proportional to $E$ indicates a non-zero nEDM. 

The frequency shift will be extracted by the spin-dependent interaction with $^3$He \cite{SpinDepCapture}. Polarized $^3$He will be injected at the time of measurement, will be subjected to a $\pi$/2 pulse to align the spins with those of the UCN, and will diffuse throughout the cell to co-habit with the neutrons. The $^3$He atomic EDM is known to be negligible compared to the neutron EDM, so the $^3$He spin can be considered to not precess under the applied electric field \cite{He3EDM}. A neutron and $^3$He in proximity and with opposing spins can react to form a proton and triton with 764 keV.  Charged decay products will cause the superfluid helium-II to scintillate, and the scintillations are observed to form the signal. This signal will oscillate in time as the neutron spins precess from aligned to anti-aligned with the $^3$He, and it will be maximized if the neutron and $^3$He spins remain in the same plane. Thus, a uniform magnetic field, particularly in the directions perpendicular to $B_{0}$ \cite{McGregor1990}, is necessary to maximize the neutron and $^3$He transverse coherence time $T_{2}^{\ast}$. 
The polarized $^3$He can also be used as a comagnetometer to measure changes in the magnetic field. SQUIDS will be used to independently monitor the precession of the $^3$He magnetization; the neutron density will be too low compared to the $^3$He to affect this signal.  The SQUID response can then be used to implement a time-dependent correction for background magnetic field shifts. 

An undesired shift in the neutron spin precession frequency arises due to the coupling of magnetic field gradients and the motional magnetic field seen by the neutron motion in an electric field, $\vec{B_{m}} = -\vec{v} \times \vec{E}/c^{2}$. This frequency shift is linearly dependent on the electric field, $E$, so it will appear as a ``false'' EDM term, $d_{f}$, in Equation \ref{eq:larmor}:

\begin{align}
\omega_{n} =  -2(\mu_{n}B_{0} \pm (d_{n} + d_{f})E)/\hbar \label{eq:larmor2}
\end{align} 

This shift $\delta\omega_{n}$ in $\omega_{n}$ due to $d_{f}$ is predicted for a rectangular cell with a linear magnetic gradient by \cite{Pignol2015}:  
\begin{align}
	\delta\omega_{n}(\omega_{0})= & -\gamma^{2}\frac{E}{c^{2}} \left( \omega_0\mathrm{Im}\left[  G_yS_{yy}\left(  \omega_0\right) +G_zS_{zz}\left(  \omega_0\right)
	\right]  + G_y\frac{L_y^2}{12}+G_z\frac{L_z^2}{12}  \right),
	\label{eq:freqshift}%
\end{align}
where $\gamma$ is the gyromagnetic ratio of the neutron, $\omega_0=\gamma B_0$ is the precession frequency of the neutrons under the holding field, $S_{yy}$ and $S_{zz}$ are the spectra of the position correlation functions found in reference \cite{Swank2016}, $G_y$ and $G_z$ are the linear magnetic field gradients, and $c$ is the speed of light. The coordinate system is defined such that the magnetic holding field is in the $x$-direction; the $y$- and $z$-directions are perpendicular to $x$, with $z$ along the axis of the cylindrical cos($\theta$) magnet. $L_{y}$ and $L_{z}$ are the lengths of the cell in those directions. Given the geometry of the SNS nEDM design, we typically require linear gradients to be $G \leq 3\times10^{-6} B_0/\mathrm{cm}$, or a false EDM $d_f \leq 1\times10^{-28} ~\mathrm{e-cm}.$
 
Novel techniques are needed to maintain magnetic uniformity in such a low-temperature environment. Use of magnetic components must be minimized, and those that are used must be placed as far from the cells as possible. 
The cells must be magnetically shielded. Additionally, temperature fluctuations in the cryogens are transferred to metal components, which generates magnetic fluctuations via Seebeck effect currents (thermoelectric effect). On the other hand, the cryogenic nature of the experiment offers the possibility of using a superconducting shield to exclude external fields via the Meissner effect. 


\subsection{$\mathbf{\nicefrac{1}{2}}$-scale magnet prototype apparatus.}   
As previously described in \cite{Adrian}, a prototype magnet has been created at $\nicefrac{1}{2}-$scale of the nEDM at SNS magnet design in order to understand the level of magnetic uniformities achievable. That work described the field uniformity of the prototype magnet as measured at room temperature. We now describe a series of upgrades to that apparatus, including a helium cryostat and a lead shield, which when cooled below 7.2 K becomes a superconductor, shielding  external magnetic fields. 

\begin{figure}[htp]
		\centering
		\includegraphics[height=5.0in]{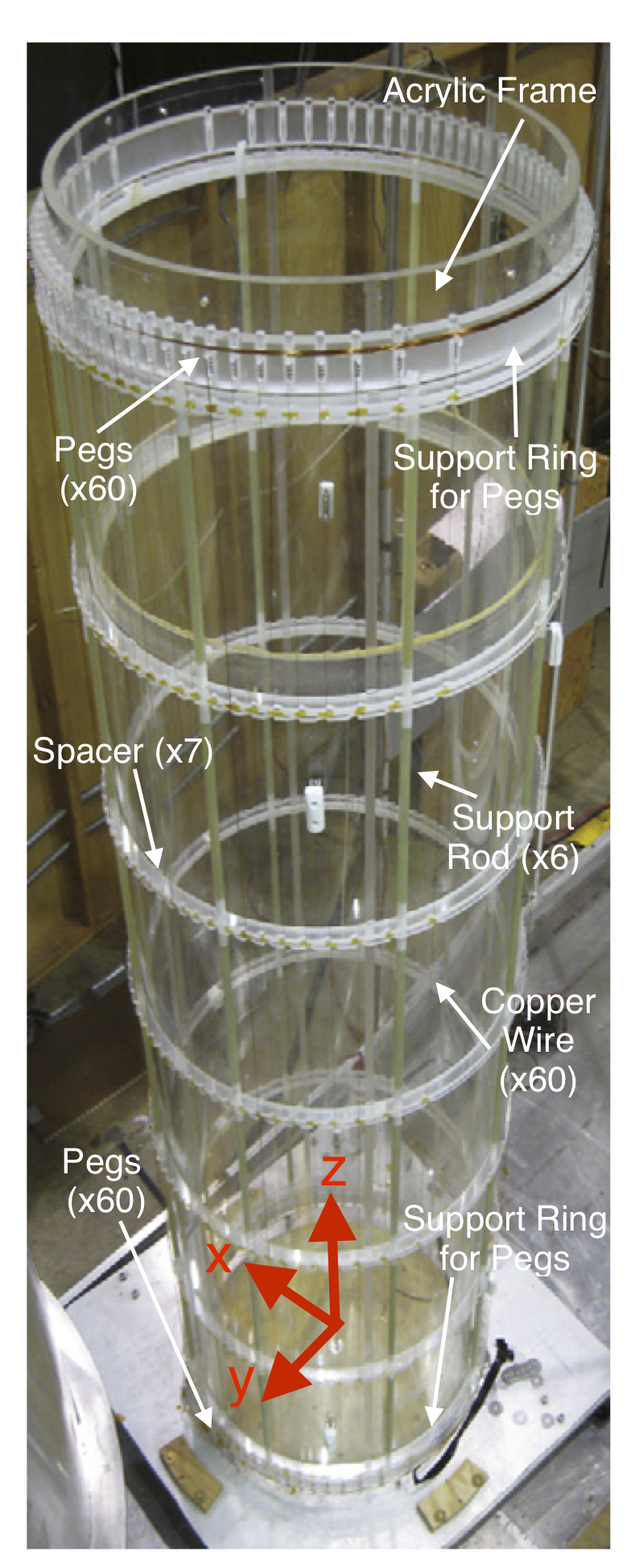}
		\caption{The $B_{0}$ coil. Copper wires are supported by pegs on springs at either end; the pegs and springs are mounted in acrylic. The frame is supported by the green/white G10/nylon rods, and seven acrylic rings have notches to keep the wires straight along the length of the cylinder.}
		\label{fig:B0}
\end{figure}

A central magnet, referred to as the ``$B_{0}$" coil and seen in Figure \ref{fig:B0}, is used to produce a uniform magnetic field, typically 3 $\mu$T, using $\sim$50 mA current. The magnet follows a ``cos($\theta$)'' design, which uses a cylindrical geometry with currents running on the surface parallel to the axis according to a cosine distribution. This design is known to generate throughout the cylinder a uniform field perpendicular to the axis of the coil \cite{JacksonCosTheta}. The center of the $B_{0}$ coil  determines the nominal origin for coordinates mentioned throughout the paper.

The $B_{0}$ coil approximates the ideal cos($\theta$) design using copper wires strung along a cylindrical surface with spacing determined by a modified cosine distribution, the details of which can be found in \cite{Adrian}. The coil used is 2.13 m tall, 0.61 m in diameter, and has 60 axial wires, which pair off into 30 coil windings. The wires are wrapped on spring-tensioned pegs to maintain straightness. The pegs are mounted in acrylic holding rings and project slightly past the rings, so the $B_{0}$ coil and trim coils sit at a diameter of 0.65 m. The holding rings are supported by vertical rods made of alternating segments of G10 and nylon, with the ratio of the lengths of the G10 and nylon segments designed so that the thermal contraction of the entire rod will match the contraction of the copper wire. Finally, the entire structure is supported on a cylindrical acrylic frame, which has vertical grooves allowing the acrylic rings to slide. 

\begin{figure}[htp]
 \centering
	\resizebox{3in}{!}{
\begin{tikzpicture}
	\newcommand*{\WireSize}{0.1cm}
        	\newcommand*{\BoRadius}{2.4cm}
	\newcommand*{\WireOffset}{0.15cm}
	\newcommand*{\NWires}{15}
	\newcommand*{\Distort}{0.005}

	\draw (0, 0) circle (\BoRadius); 

	\draw[thick, dotted] (-0.185cm, 0.185cm) rectangle (0.185cm, 0.465cm); 
	\draw[thick, dotted] (-0.185cm, -0.185cm) rectangle (0.185cm, -0.465cm); 

	\draw[->] (-1.0cm,-1.5cm) -- (-1.0cm,1.5cm);
	\draw[->] (0cm,-1.5cm) -- (0cm,1.5cm)
	node [left,text width=1cm,align=center,midway]
	    {
      		$B_{0}$
    	    };
	\draw[->] (1.0cm,-1.5cm) -- (1.0cm,1.5cm);

	\draw[->] (4.0cm, 0cm) -- (4.0cm, 0.5cm) 
	node [above]
	    {
      		$x$
    	    };
	\draw[->] (4.0cm, 0cm) -- (3.5cm, 0cm) 
	node [left]
	    {
      		$y$
    	    };
	\draw (4.0, 0cm) circle (0.15cm); 
	\filldraw[fill=black](4.0, 0cm) circle (0.05cm) 
	node [below right=0.075cm]
	    {
      		$z$
    	    };
	
	\foreach \myangle in {0, 1, 2,...,14}
		\filldraw[fill=black]({atan2( (1-\Distort)*(1-\Distort), sqrt(1/(((0.5+\myangle)/\NWires)*((0.5+\myangle)/\NWires)) -1) )}:
			\BoRadius + \WireOffset + \WireSize/2) circle(\WireSize/2.0);	
	\foreach \myangle in {0, 1, 2,...,14}
		\filldraw[fill=black](-{atan2( (1-\Distort)*(1-\Distort), sqrt(1/(((0.5+\myangle)/\NWires)*((0.5+\myangle)/\NWires)) -1) )}:
			\BoRadius + \WireOffset + \WireSize/2) circle(\WireSize/2.0);	\foreach \myangle in {0, 1, 2,...,14}
	\foreach \myangle in {0, 1, 2,...,14}
		\filldraw[fill=black]({180-atan2( (1-\Distort)*(1-\Distort), sqrt(1/(((0.5+\myangle)/\NWires)*((0.5+\myangle)/\NWires)) -1) )}:
			\BoRadius + \WireOffset + \WireSize/2) circle(\WireSize/2.0);	
	\foreach \myangle in {0, 1, 2,...,14}
		\filldraw[fill=black]({180+atan2( (1-\Distort)*(1-\Distort), sqrt(1/(((0.5+\myangle)/\NWires)*((0.5+\myangle)/\NWires)) -1) )}:
			\BoRadius + \WireOffset + \WireSize/2) circle(\WireSize/2.0);	
			
	\foreach \myangle in {3, 8, 13}
		\filldraw[fill = red, draw=red]({atan2( (1-\Distort)*(1-\Distort), sqrt(1/(((0.5+\myangle)/\NWires)*((0.5+\myangle)/\NWires)) -1) )}:
			\BoRadius*1.015 + \WireOffset + \WireSize/2 + \WireSize*2) rectangle ++(\WireSize,\WireSize);
	\foreach \myangle in {3, 8, 13}
		\filldraw[fill = red, draw=red](-{atan2( (1-\Distort)*(1-\Distort), sqrt(1/(((0.5+\myangle)/\NWires)*((0.5+\myangle)/\NWires)) -1) )}:
			\BoRadius*1.015 + \WireOffset + \WireSize/2 + \WireSize*2) rectangle ++(\WireSize,\WireSize);
	\foreach \myangle in {3, 8, 13}
		\filldraw[fill=red, draw=red]({180-atan2( (1-\Distort)*(1-\Distort), sqrt(1/(((0.5+\myangle)/\NWires)*((0.5+\myangle)/\NWires)) -1) )}:
			\BoRadius*1.015 + \WireOffset + \WireSize/2 + \WireSize*2) rectangle ++(\WireSize,\WireSize);
	\foreach \myangle in {3, 8, 13}
		\filldraw[fill=red, draw=red]({180+atan2( (1-\Distort)*(1-\Distort), sqrt(1/(((0.5+\myangle)/\NWires)*((0.5+\myangle)/\NWires)) -1) )}:
			\BoRadius*1.015 + \WireOffset + \WireSize/2 + \WireSize*2) rectangle ++(\WireSize,\WireSize);
			
	\filldraw[fill=blue, draw=blue] ({atan2( (1-\Distort)*(1-\Distort), sqrt(1/(((0.5+14)/\NWires)*((0.5+14)/\NWires)) -1) )}:
			\BoRadius*1.015 + \WireOffset + \WireSize/2 + \WireSize*2) ellipse(\WireSize/2.0 and \WireSize)	
	node [above = 0.1 cm]
	    {
      		\small x5
    	    };

	\filldraw[fill=blue, draw=blue](-{atan2( (1-\Distort)*(1-\Distort), sqrt(1/(((0.5+14)/\NWires)*((0.5+14)/\NWires)) -1) )}:
			\BoRadius*1.015 + \WireOffset + \WireSize/2 + \WireSize*2) ellipse(\WireSize/2.0 and \WireSize)
	node [below = 0.1 cm]
	    {
      		\small x5
    	    };	
	\filldraw[fill=blue, draw=blue]({180-atan2( (1-\Distort)*(1-\Distort), sqrt(1/(((0.5+14)/\NWires)*((0.5+14)/\NWires)) -1) )}:
			\BoRadius*1.015 + \WireOffset + \WireSize/2 + \WireSize*2) ellipse(\WireSize/2.0 and \WireSize)
	node [above = 0.1 cm]
	    {
      		\small x5
    	    };
	\filldraw[fill=blue, draw=blue]({180+atan2( (1-\Distort)*(1-\Distort), sqrt(1/(((0.5+14)/\NWires)*((0.5+14)/\NWires)) -1) )}:
			\BoRadius*1.015 + \WireOffset + \WireSize/2 + \WireSize*2) ellipse(\WireSize/2.0 and \WireSize)
	node [below = 0.1 cm]
	    {
      		\small x5
    	    };
			

\end{tikzpicture}		
}
\caption{Schematic top view of the $B_{0}$ coil, showing the locations of the $B_{0}$ wires and trim wires. Black dots and red squares indicate the $B_{0}$ wires and its trim coil, respectively. Blue ellipses indicate the rectangular quadratic trim coil, which is wrapped 5 times. The thin black circle represents the acrylic $B_{0}$ frame, which sits at 0.61 m OD. All the wires are wrapped around pegs at a diameter of 0.65 m. The fiducial measurement region is indicated in dotted lines in the center. One region extends from 2.5 cm to 6.3 cm in x, -2.5 cm to 2.5 cm in y, and -10 cm to 10 cm in z (into and out of the page); the other region is the same but with negative x coordinates. }
	\label{fig:B0Schematic}
\end{figure}
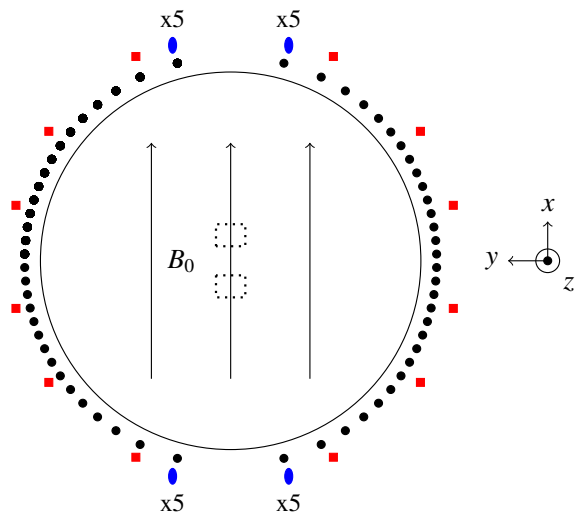

Trim coils are used to compensate for field non-uniformities arising from local magnetic impurities or from the fact that the coil is discrete and not an ideal cos($\theta$) design. It was found that three pairs of trim coils were necessary to cancel ambient gradients, as shown schematically in Figure \ref{fig:B0Schematic}: a cos($\theta$) coil with 12 wires (6 windings) to trim the field in the $B_{0}$ direction; a rectangular coil with 4 wires (2 windings) to trim quadratic fields in the $B_{0}$ direction, wrapped 5 times to increase its magnitude; and circumferential coils around the $B_{0}$ frame to control the field in the vertical direction. Each coil was split about its plane of symmetry, so that the current in each half could be controlled separately. This allowed for switching between a mode of maximum uniformity (for field magnitude control) or a gradient control mode. The trim coils are controlled by LabVIEW and powered by a Measurement Computing 16-bit digital to analog converter (DAC) (USB-3114) with $\pm$40 mA current capability per channel \cite{Labview, MC}. 

\begin{figure}[htp]
		\centering
		\includegraphics[width=\textwidth, keepaspectratio]{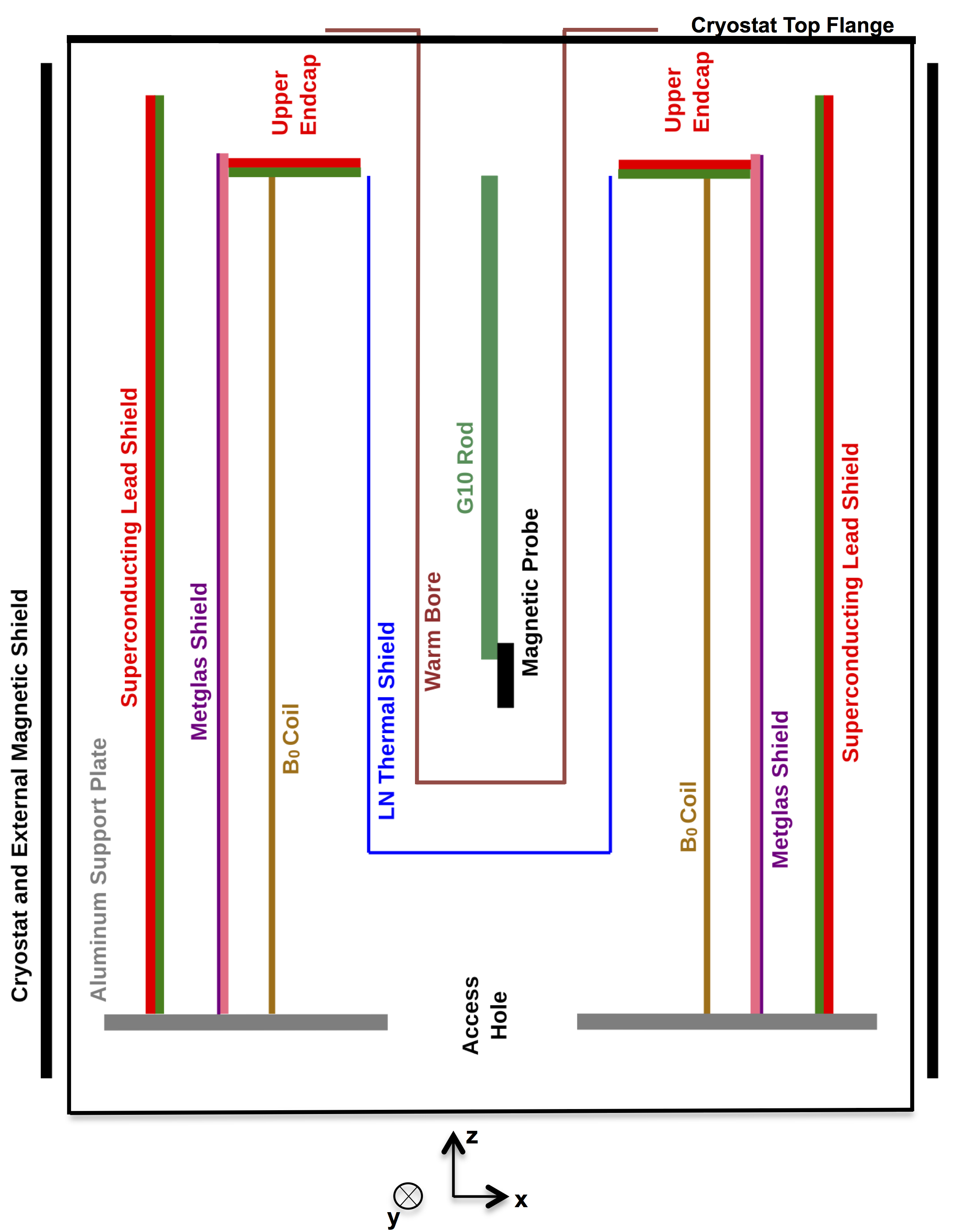}
		\caption{A schematic showing the nested arrangement of coils and magnetic and thermal shielding. Additional thermal shielding between the cryostat top flange and the rest of the components is not shown. The coordinate origin with $x = y = z = 0$ is at the center of the $B_{0}$ coil. Not to scale.}
		\label{fig:HalfScaleSchematic}
\end{figure}

The magnet system includes magnetic shielding to improve uniformity. 
To accommodate the cryogenic superconductor, the package of magnetic shielding and coils 
is operated inside a cryostat, discussed below. The entire apparatus is shown schematically in Figure \ref{fig:HalfScaleSchematic}. The cryostat itself is wrapped in four layers of Metglas 2705M and one layer of mu-metal to reduce the size of environmental fields, primarily the Earth's field. The Earth's unshielded magnetic field is of magnitude $\sim$
50 $\mu$T at the location of the cryostat, primarily in the vertical direction, along the axis of the apparatus. 

\begin{figure}[htp]
		\centering
		\includegraphics[height=3.0in]{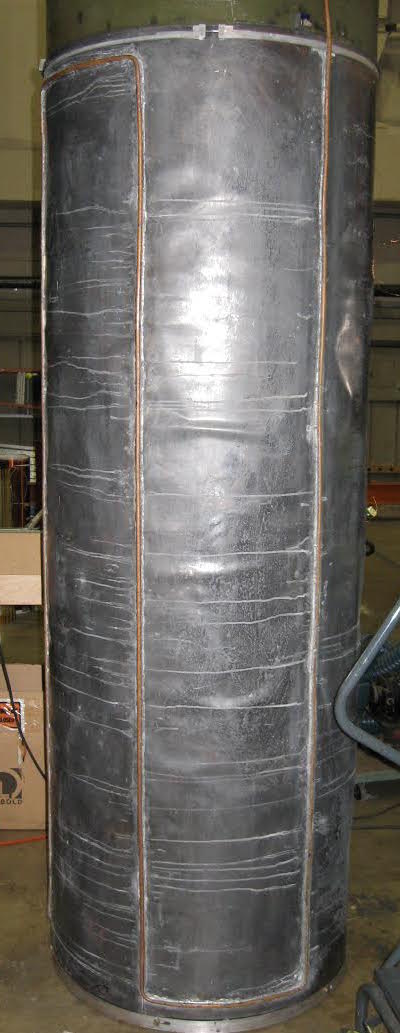}
		\caption{Lead shield used for superconducting magnetic shielding. The lead, 2.13 m high, is mounted on a G10 cylinder. Soldered copper cooling lines allow for thermal contact with liquid helium.}
		\label{fig:LeadShield}
\end{figure}

Inside the cryostat, the outermost layer of the magnet package is a cylindrical superconducting lead shield, shown in Figure \ref{fig:LeadShield}. The lead is 0.75 m in diameter, 2.2 m long, and 0.8 mm thick, and it is supported by a G10 fiberglass frame. G10 is chosen for its rigidity and its low thermal contraction coefficient. Copper pipes soldered to the lead and taped to the frame carry liquid helium to cool the shield. An ideal superconducting shield would expel all external magnetic flux via the Meissner effect, but in practice, magnetic flux can be trapped in impurities or inhomogeneities in the lead as it transitions below the superconducting transition temperature, $T_{C} = 7.2$K. Thus, it is important to minimize the magnetic fields at the surface of the lead before cooling it below the superconducting transition. Below $T_{C}$, the lead is very effective at mitigating time-varying external fields. 

The $B_{0}$ field is perpendicular to the axial lead shield, but a superconductor admits no perpendicular magnetic field at its surface. Thus, there is a field distortion at the lead, and it is necessary to provide a flux return with the correct boundary conditions. To this end, a ``Metglas shield'' was placed inside the lead shield, consisting of six layers of Metglas 2705M \cite{Metglas}, wrapped on a Vylon tube \cite{Vylon} 2.25 m in hight and 0.72 m diameter. The layers are composed of strips 22 $\mu$m thick and 5.08 cm wide, with alternate longitudinal and circumferential layers to allow for magnetic flux continuity in both directions along the cylindrical surface. Metglas 2705M was chosen as it retains high permeability from room temperature to cryogenic temperatures. The Metglas shield also provides magnetic shielding \cite{RTMetglasStudies} from environmental fields.

The mu-metal/Metglas external shield and the internal Metglas shield both include degaussing coils. The coils consist of 20 turns of 0.81 mm wire wrapped along the inner and outer surfaces of the shield, parallel to the axis. The loops penetrate both faces of each support cylinder, so as to form a segment of a toroid around each shield. The coils are powered by a California Instruments 801RP power supply, controlled via LabVIEW \cite{CaliforniaInst}. The degaussing current is a 60 Hz sinusoid modulated to ramp from zero amplitude to a peak of 5.5 Amperes, and then drop to zero again, in $\sim$5 min. The external and internal shields are degaussed sequentially, in that order. 

\begin{figure}[htp]
		\centering
		\includegraphics[width=\textwidth, keepaspectratio]{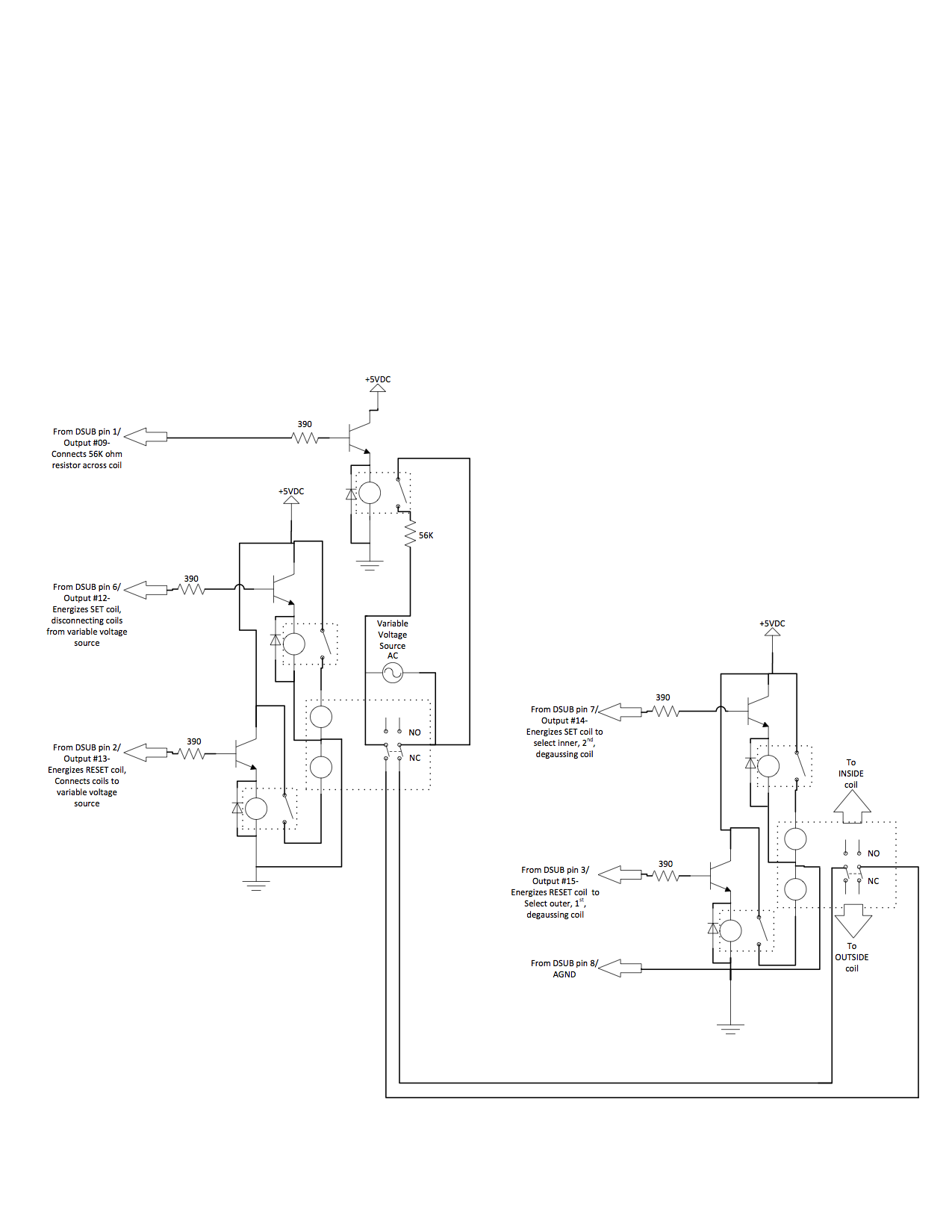}
		\caption{Circuit used to control mu-metal and Metglas degaussing. A 5.5 A current sweep is provided by the variable voltage source. One switch decouples the source from the degauss coils to protect the coils from noisy bursts of current at the end of the sweep. Another switch alternates between degaussing the external and internal magnetic shields. }
		\label{fig:degausscircuit}
\end{figure}

We noticed that the 801RP supply produced a small burst of uncontrolled current at the end of the ramp down, leaving a small field trapped in the internal Metglas. To eliminate this, we installed a transformer between the supply and the coils to isolate the coils. We also installed a LabVIEW-controlled relay between the coils and the power supply which opens the circuit just before the burst occurs. A circuit diagram for this relay system is shown in Figure \ref{fig:degausscircuit}.

\begin{figure}[htp]
		\centering
		\includegraphics[width=3.0in]{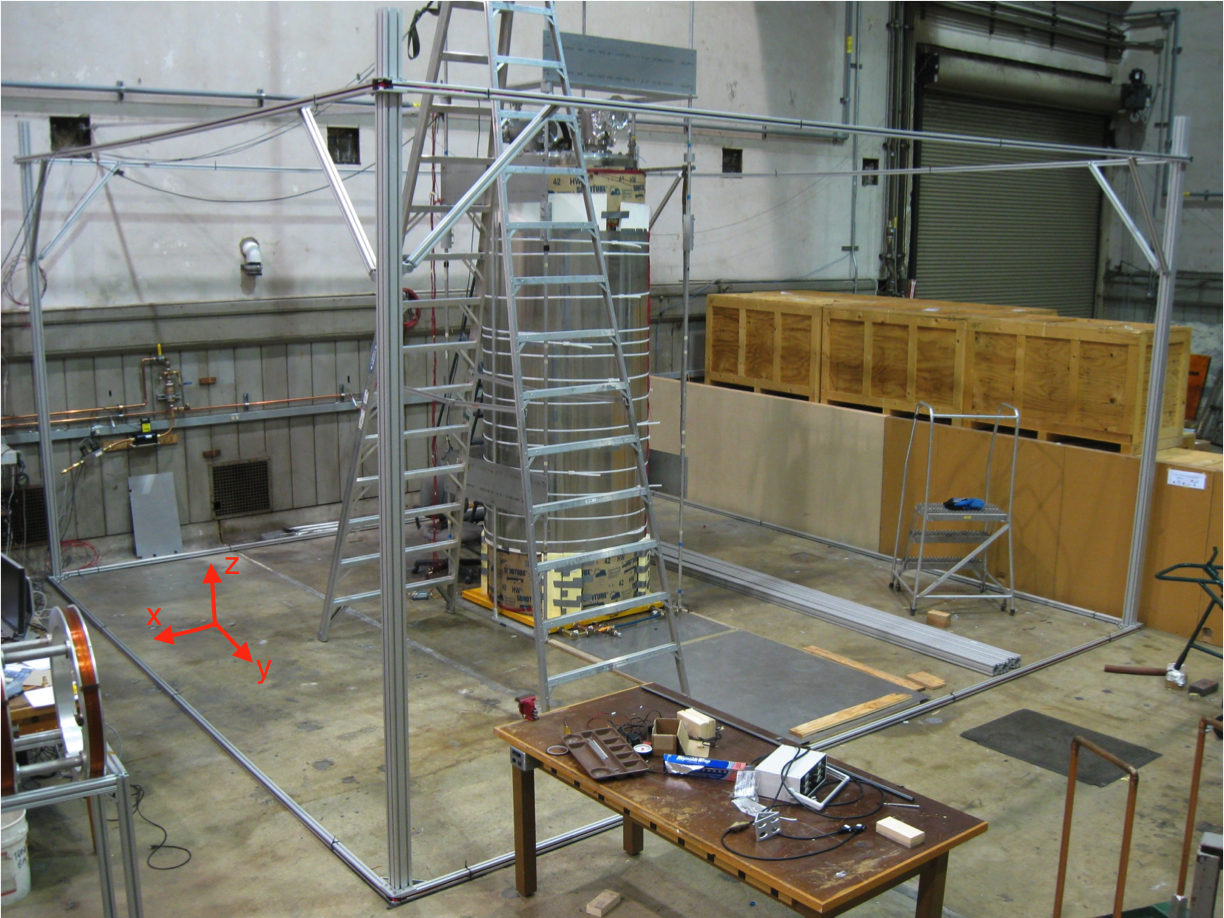}
		\caption{Ambient field compensation coils are wound on the metal bars shown surrounding the cryostat. Additional fine gradient coils are not shown.}
		\label{fig:AmbientCancelling}
\end{figure}

To further mitigate the flux trapped in the lead shield due to the Earth's field, rectangular compensation coils are installed in all three directions surrounding the magnetic testing area; see Figure \ref{fig:AmbientCancelling}. The coils producing field in the z-direction are 6.1 m x 6.1 m with 3.7 m separation, while the x-coils and y-coils are 6.1 m x 3.7 m with 6.1 m separation. Each direction has a main and trim coil, respectively to adjust magnitude and gradients. Additional coils are installed in the x-direction that bisect the main coil, effectively forming four equal coils of half the height and width; this provides finer gradient control. Coils are simply tuned by hand after the cryostat and magnetic package are installed. The field at the center of the $B_{0}$ coil is monitored in three orthogonal directions and each component is minimized.

The magnetic field is measured with a 3-axis, low-noise fluxgate magnetometer (Bartington Mag-03MSESL100) \cite{Bartington}. The magnetometer signal is read out with a Signal Conditioning Unit (Bartington SCU1) and passed to a LabVIEW monitoring system via a 24-bit analog to digital converter (ADC) (Measurement Computing USB-2408). For automated field mapping, the probe is coupled to a stepper motor via a 3 m long G10 rod. The motor sits on top of an aluminum stand, placing any magnetic fields it produces about 3 m from the field mapping region. A 3-axis National Instruments motion controller (MID-7604) allows LabVIEW control of the motor \cite{NI}. 	

\subsection{Superconducting endcap}
\label{subsec:Endcap}

As part of experimental design optimization, it is of interest for the SNS nEDM experiment to explore the possibility of closing the ends of the superconducting lead shield. This would provide improved uniformity of the $B_{0}$ field, as well as additional shielding from external magnetic fields. Fringing fields from the open ends would also be reduced, so designs with more compact magnets and shields would become available, reducing engineering costs and challenges. To allow assembly of internal components, the ends would have to be removable; thus, a lead disk, or ``endcap'', is required to cover the ends of the cylindrical portion of the lead shield and form a nearly hermetic shield. 

\begin{figure}[htp]
 \centering
   \begin{tabular}{cc}  
   \includegraphics[width={0.5\textwidth}]{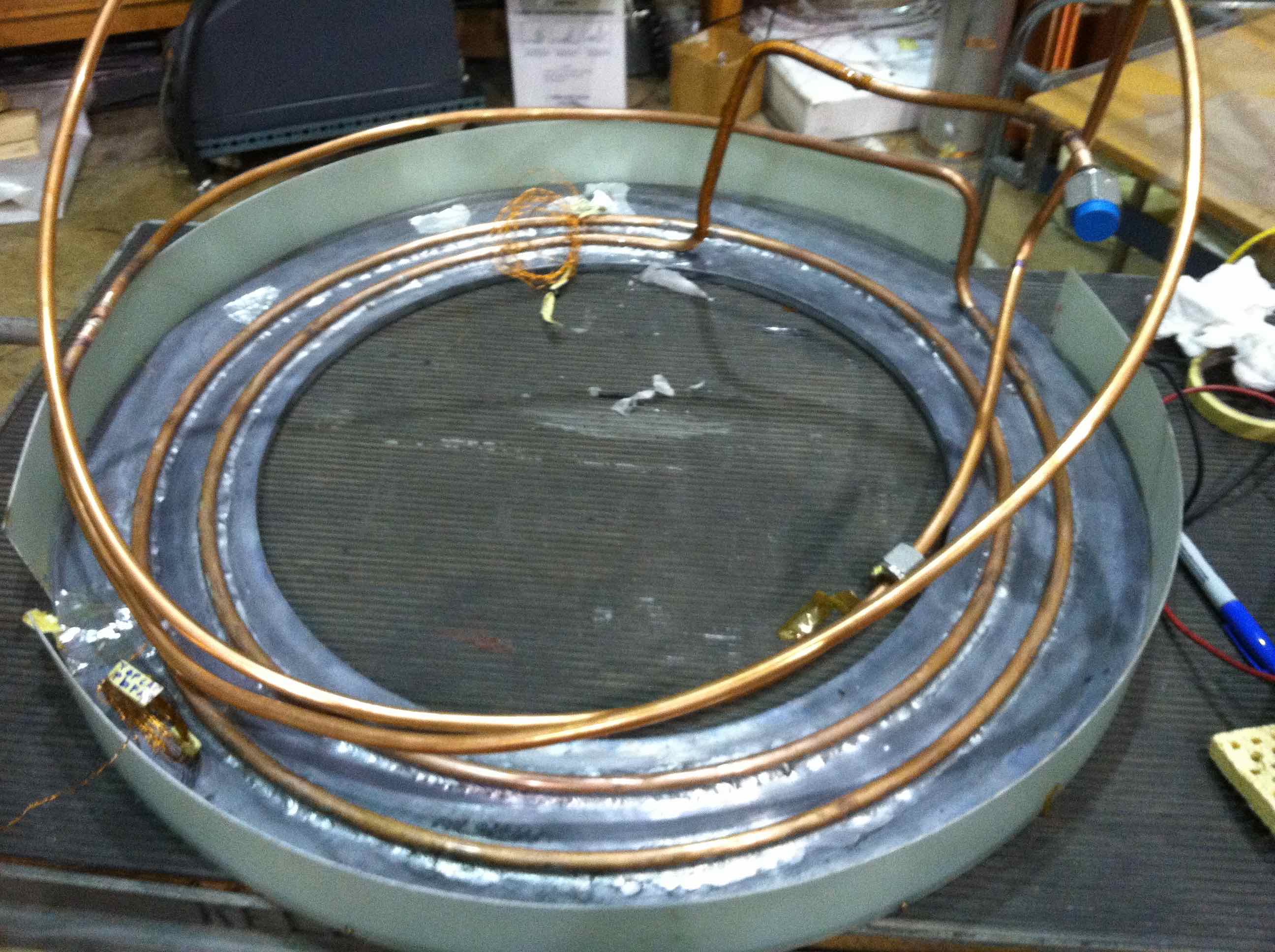}&
   \includegraphics[width={0.5\textwidth}]{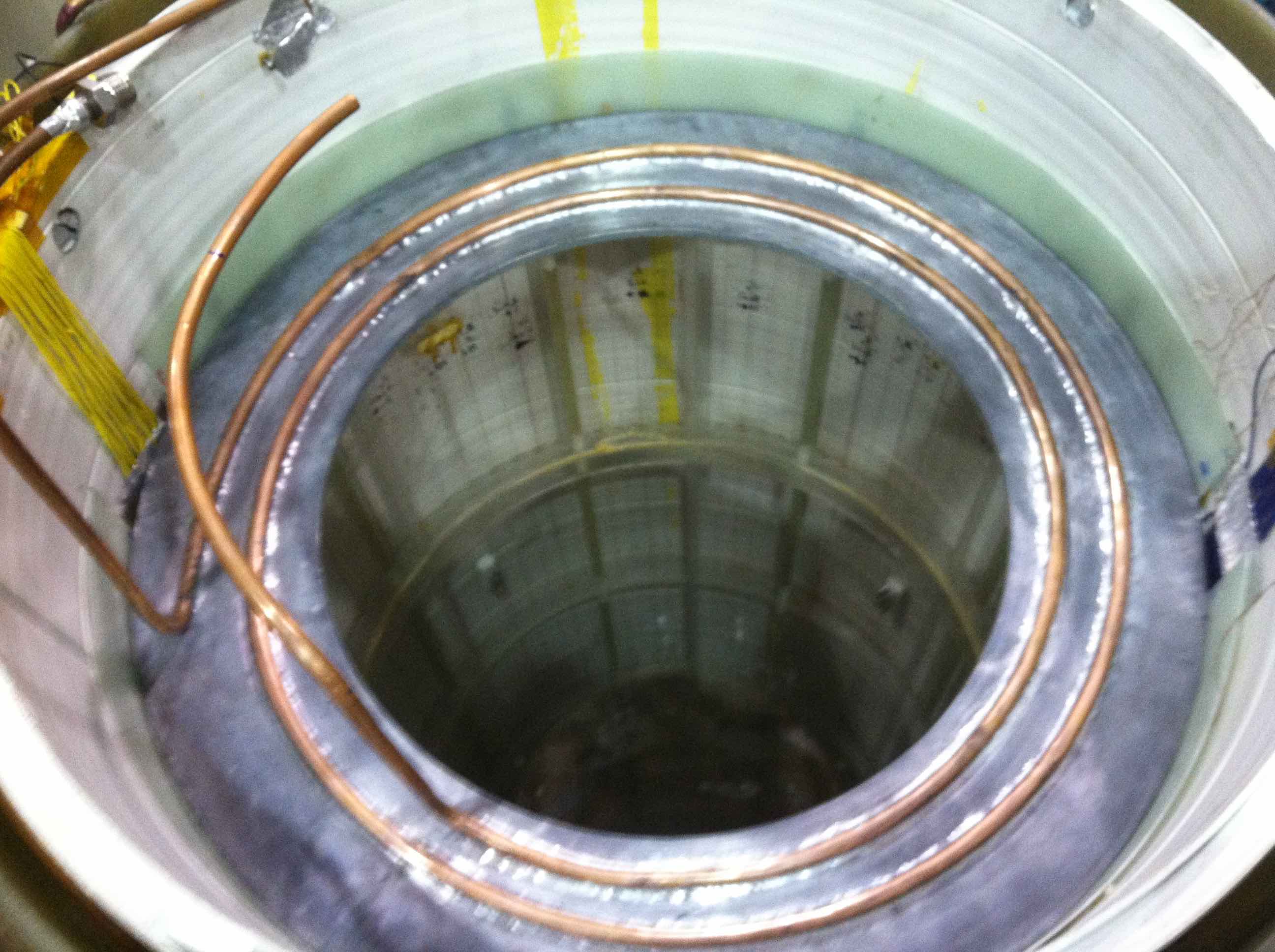}\\  
    \end{tabular}
  \caption{Superconducting lead endcap used to partially close an open end of the superconducting shield. The lead is supported on a G10 frame (not seen, underneath) and has a G10 wall to retain its position. Copper tubing soldered to the lead provides liquid helium cooling. On the left, the endcap is pictured prior to installation, viewed from above, with fittings installed and heater wire wrapped around the input. On the right, the endcap is shown in its installed position, which is resting on top of $B_{0}$ coil and inside the white Vylon support for the internal Metglas shield. Part of the $B_{0}$ coil can be seen through the hole in the endcap.} 
     \label{fig:EndcapFig}
\end{figure}

We created such a removable lead disk to fit inside the internal Metglas shield, roughly at the height of the top of the lead shield; see Figures \ref{fig:HalfScaleSchematic} and \ref{fig:EndcapFig}. To accommodate the warm bore, the disk is annular, with OD 66 cm and ID 43.2 cm. It sits on a G10 disk for support, which rests on top of the $B_{0}$ magnet frame. The lead has independent soldered-copper cooling lines, and the entrance to the lines was wrapped with heater wire to allow the superconducting state to be controlled for comparison. 

There are significant limitations to this endcap arrangement. Only an upper endcap was installed; a lower endcap would have required modifications of the lower aluminum mount. This, combined with the central hole in the endcap, means the lead shield is not completely hermetic in this arrangement. Furthermore, the length of the lead shield means that the endcap is too far away to measurably affect the field in the region of interest inside the $B_{0}$ coil. Thus, this prototype cannot be used to assess the endcap's effect on magnetic gradients; this will instead be explored in the future. Nevertheless, this prototype can be used to validate simulations of endcap designs, which improves the reliability of simulations for future designs. Even with a central hole, simulations of the annular endcap design predict a measurable improvement on field uniformity in the vicinity of the endcap. These results also are presently being used to further optimize the magnet package for the experiment. 

\subsection{Cryogenic Apparatus}
\label{sec:CryoApp}
The package of $B_{0}$ magnet and shields is operated in a 3 m tall, 1 m diameter triple-walled stainless-steel cryostat. The cryostat is liquid nitrogen and liquid helium cooled, and contains bayonet penetrations for liquid helium service to the package. The lid contains a deep indent in the center, a ``warm bore'' $\sim$1.5 m deep and 0.28 m diameter, to allow a room-temperature probe to access to the interior of the $B_{0}$ coil.

Inside the cryostat, the magnet package is supported from below by a plate of 1100-alloy aluminum, which is suspended from the cryostat lid by G10 rods. This alloy was chosen for having a larger thermal conductivity than other aluminum alloys at cryogenic temperatures. The plate is helium cooled, providing additional thermal shielding and conductive cooling. A helium-cooled aluminum radiation shield (``mushroom") is placed on top of the magnet package, covering the top of the package and part of its side. The mushroom has a cut-out in the top for the warm bore to pass through. Both the aluminum plate and the mushroom have 3003-alloy aluminum cooling lines attached via dip-brazing \cite{DipBraze}.\footnote{Initially, the plate and mushroom were copper-plated and had copper cooling lines attached by brazing. However, the plating process included nickel, leaving measurable magnetization. The copper plating and copper tubing were removed by abrasion and chemical processes before attaching the aluminum cooling lines.} The aluminum cooling lines are coupled to stainless-steel penetrations in the cryostat lid using friction-welded \cite{FrictionWeld} aluminum-to-stainless steel VCR $\textregistered$ connectors \cite{VCR}. The room-temperature bore is surrounded by an aluminum thermal shield 0.33 m in diameter, which is cooled via copper tubes taped to the sides. The warm bore and radiation shields are covered in 10-layer multi-layer insulation (MLI) to improve thermal shielding. 

\subsection{Cryogenic Design}
\label{sec:CryoDesign}

Electrical currents can be generated by thermal gradients in the metal components via the Seebeck effect. These currents produce stray magnetic fields that can add magnetic noise and non-uniformities in the measurement region. Aluminum was preferred for metal parts inside the cryostat since it has a lower Seebeck coefficient than other natural choices, such as copper. Still, magnetic fluctuations correlated with temperature gradients were observed, possibly originating in the copper cooling lines or in the aluminum warm bore. Hence, it was necessary to implement measures to stabilize the temperatures in the liquid cryogen.

\begin{figure}[htp]
		\centering
		\includegraphics[width=5.25in]{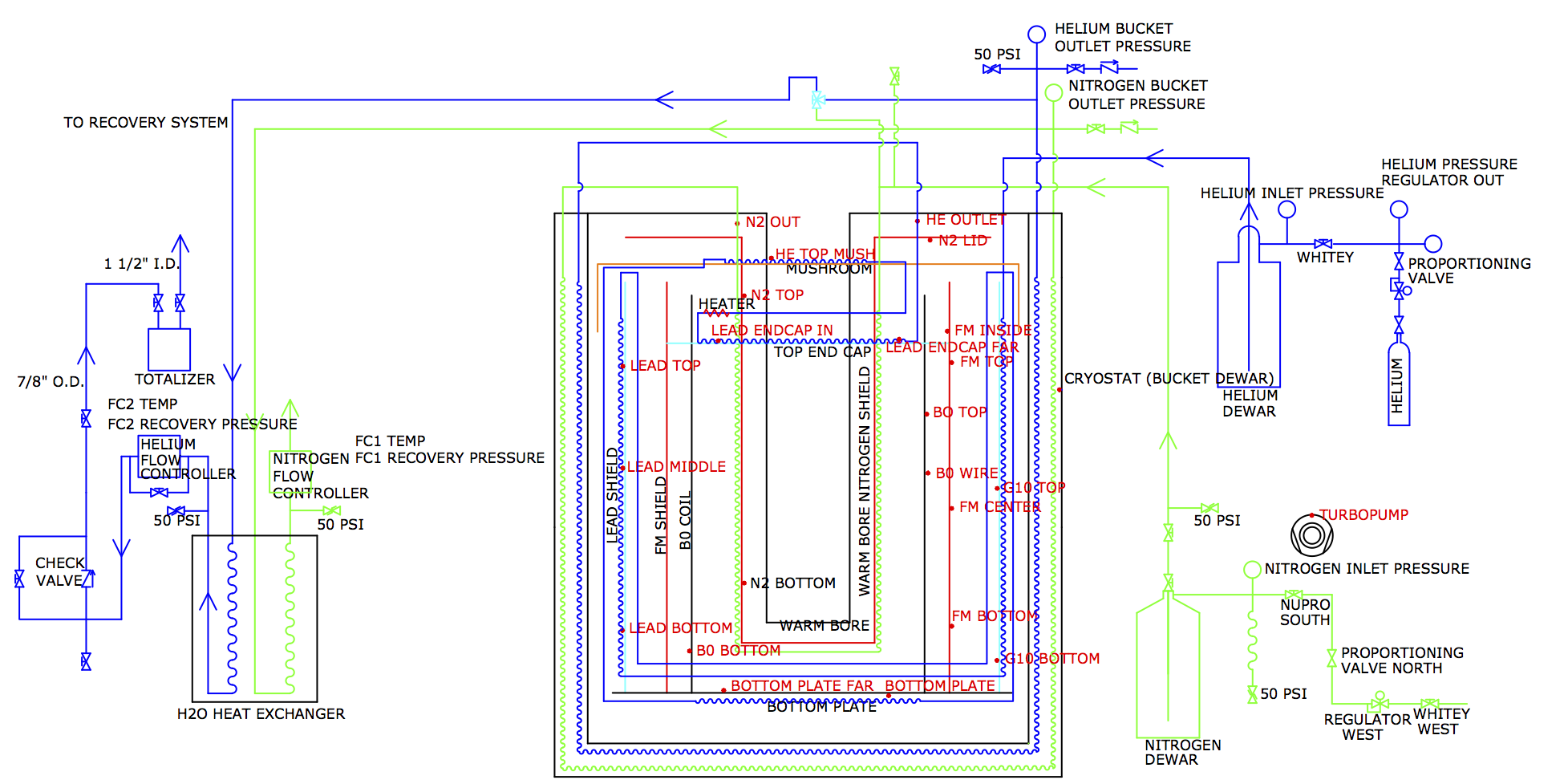}
		\caption{A schematic of liquid cryogen control system. Green and blue lines represent nitrogen and helium cooled systems, respectively, and wavy lines indicate heat exchange coils. The 3-way valve, shown in teal, allows pre-cooling of the helium system with nitrogen. Red dots indicate temperature sensor locations. Liquid cryogen flow proceeds in the direction of the arrows. Gas cylinders pressurize liquid cryogens at the input, and the flow rates are regulated at the output. A liquid-water bath (``H$_{2}$O heat exchanger'') protects the flow controllers from low temperatures. }
		\label{fig:CryoSys}
\end{figure}

A schematic of the cryogenic flow system is shown in Figure \ref{fig:CryoSys}. Liquid nitrogen and helium supply dewars are pressurized with the respective gas to increase flow and ensure uniformity of the supply pressure. The pressure is supplied through regulators which are controlled using LabVIEW via a 16-bit digital to analog converter (DAC) (Measurement Computing USC-3102). Mass flow controllers (Alicat MCR series, \cite{Alicat}) on the cryogen exhaust dramatically improve the temperature and magnetic stability of the system. The controller for the helium is modified, at our request of the manufacturer, with an option to control on either pressure or mass flow as the process variable. Pressure control was found to further increase temperature stability. The flow controller parameters are set using Labview via serial connections. 

Pressure and temperatures sensors are located in key places in the system, shown in Figure \ref{fig:CryoSys}. The sensors are monitored in real time with a Labview-based ``slow-control'' program, the same program that controls the cryogen regulators and flow controllers. Pressure data is read in using a 12-bit National Instruments ADC (USB-6008). Temperature data is acquired from a mixture of $\sim$30 platinum resistance temperature detectors and Si diodes, respectively for liquid-nitrogen and liquid-helium cooled areas of the system. These signals are read and calibrated using control units (Lakeshore 218 \cite{Lakeshore}), and from there are passed into slow-control via serial connection. 

\section{Results}
\label{sec:Results}

\subsection{Thermal Stability}
\label{sec:ResultsTherm}

\begin{figure}[htp]
	\centering
	\includegraphics[width=5.0in]{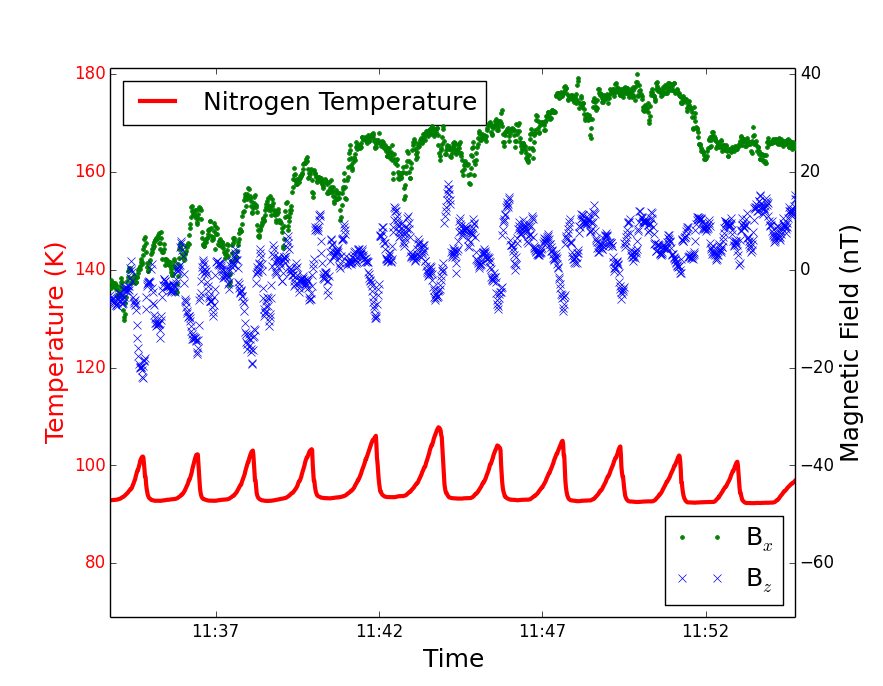}
	\caption{Magnetic field fluctuations correlated with temperature fluctuations in the liquid-nitrogen-cooled portions of the thermal shielding, as measured at N$_{2}$ Bottom (see Fig. \ref{fig:CryoSys}). $B_{z}$ oscillates in phase with the temperature, while $B_{x}$ oscillates with the same period, but out of phase. The liquid nitrogen mass flow rate is regulated. Stopping the liquid-nitrogen flow reduced the magnetic fluctuations to negligible levels (not shown). The magnetic field is measured at the center of the $B_{0}$ coil. }
	\label{fig:LNFluct}
\end{figure}

Fluctuations in cryogenic component temperatures were observed to correlate strongly with magnetic fluctuations in both nitrogen-cooled and helium-cooled components. Figure \ref{fig:LNFluct} shows this correlation for the temperature of the warm bore liquid nitrogen thermal shielding (N$_{2}$ Bottom in Figure \ref{fig:CryoSys}). Here the mass flow rate of the nitrogen vapor is controlled. To mitigate these fluctuations in liquid nitrogen, the flow was simply shut off during sensitive magnetic measurements. Temperatures in the nitrogen-cooled components were allowed to drift upward, which occurred at a rate of 1 K/hr. The magnetic field was measured at the center of the $B_{0}$ coil.



\begin{figure}[htp]
	\centering
	\includegraphics[width=5.0in]{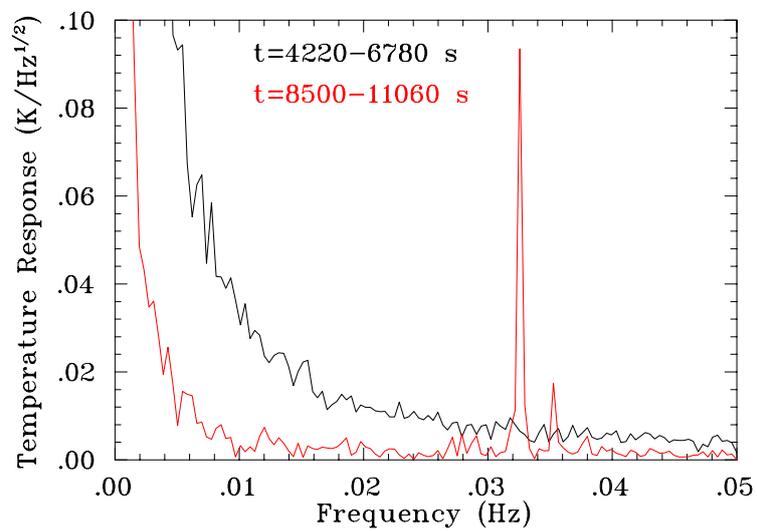}
	\caption{Fourier transform of the time-dependence of the lead superconductor temperature (``Lead Top'' in Fig. \ref{fig:CryoSys}) with liquid-helium flow controlled based on output pressure (black) and mass flow rate (red). Pressure control eliminates the strong characteristic thermal fluctuation at 0.032 Hz. The time intervals in the legend are sequential; the control mode was changed in the intervening time to demonstrate the effect. Note that the liquid-nitrogen flow was stopped for these measurements.}
	\label{fig:PbTempFFT}
\end{figure}

\begin{figure}[htp]
	\centering
	\includegraphics[width=5.0in]{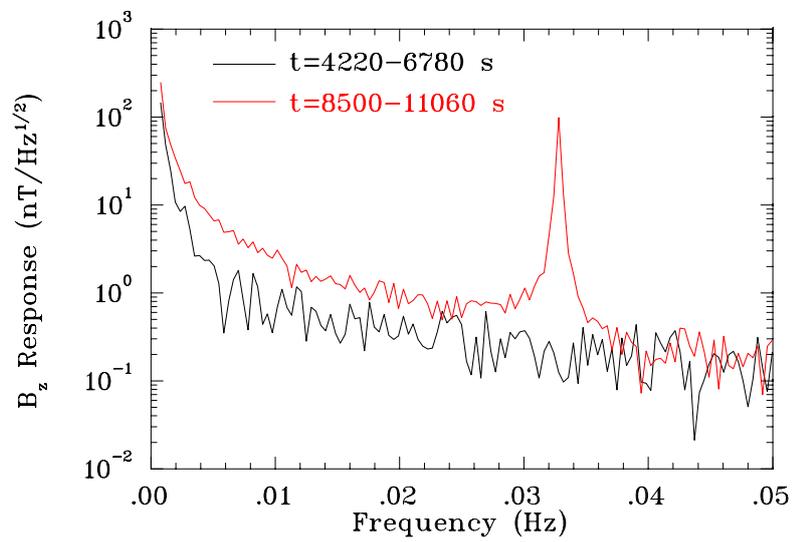}
	\caption{Fourier transform of the time-dependence of the vertical component of the magnetic field measured at the center of the $B_{0}$ magnet. The time intervals correspond to Figure \ref{fig:PbTempFFT}, with liquid helium flow controlled based on output pressure (black) and mass flow rate (red). The magnetic field suffers a fluctuation precisely corresponding to that of the temperature when in mass-flow-control mode, and is correspondingly eliminated in pressure-control mode. Note that the liquid-nitrogen flow was stopped for these measurements.}
	\label{fig:BzFFT}
\end{figure}

\begin{figure}[htp]
	\centering
	\includegraphics[width=5.0in]{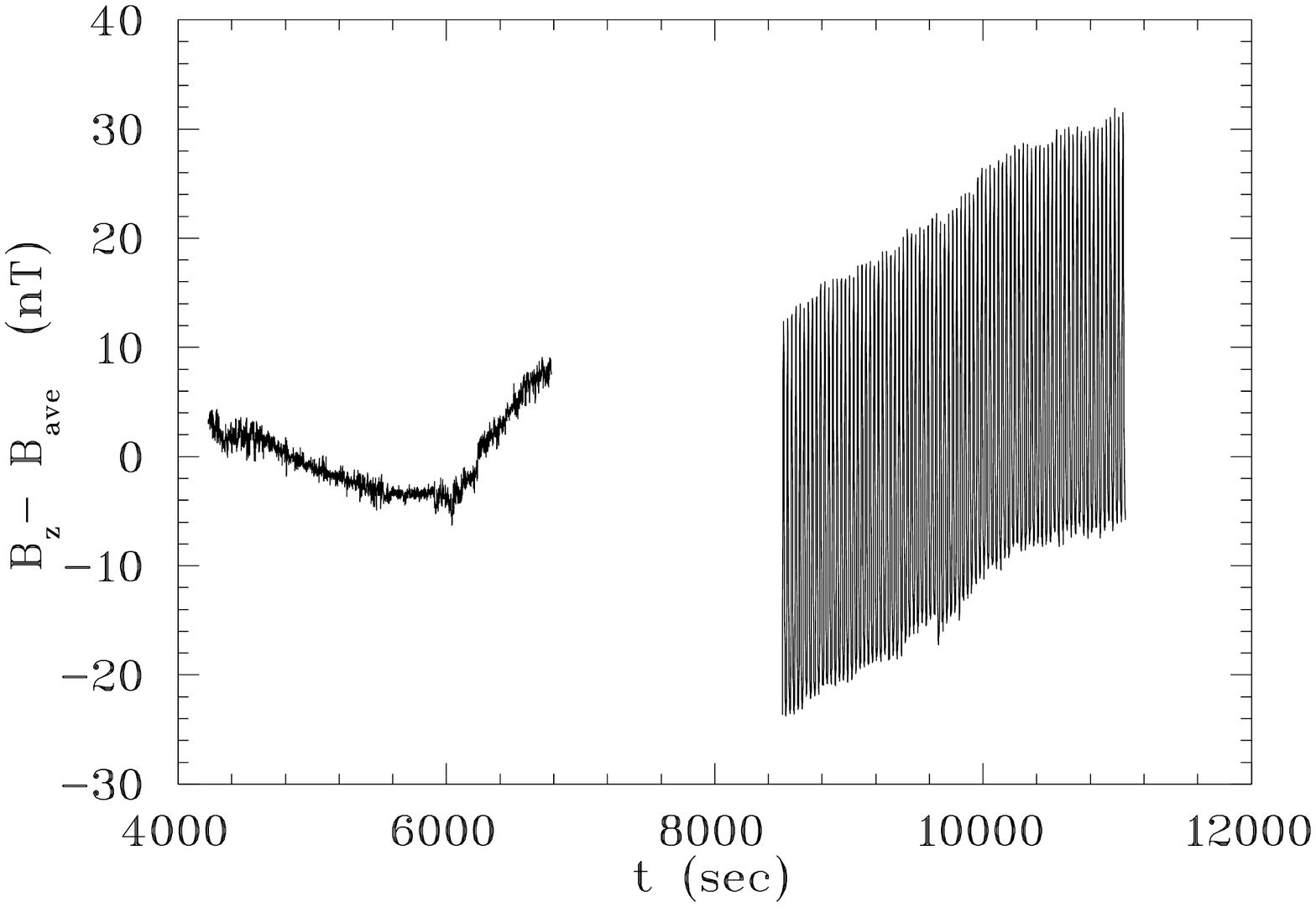}
	\caption{Deviation of $B_{z}$ from average, with liquid helium flow pressure-controlled before 8000 s and mass-flow-controlled after 8000 s. The control mode was changed to demonstrate the effect on short-term fluctuations. Note that the warm bore thermal shield is not cooled during this time, so as to reduce noise. However, this allows the temperature of the shield to drift, leading to corresponding drifts in the magnetic field on timescales of $\sim$ 1000 s. The magnitude of the observed drifts is not reduced by magnetic shielding, since the fluxgate is not magnetically shielded from the warm bore.}
	\label{fig:Bz}
\end{figure}



The corresponding correlations between the fluctuations in the liquid helium temperature and measured magnetic field are shown in Figures \ref{fig:PbTempFFT} and \ref{fig:BzFFT} for mass flow control. A 0.032 Hz characteristic temperature oscillation frequency is clearly present when the helium mass flow rate is controlled. The helium temperature is measured in the lead shield using the sensor labeled ``Lead Top'' in Fig. \ref{fig:CryoSys}. For precise measurements of the magnetic field it was necessary to eliminate this oscillation, and shutting off the helium flow was not possible as the superconducting lead would rise above the critical temperature too quickly. In order to reduce the thermal fluctuations, we changed from mass flow control to vapor pressure control by replacing the liquid helium mass flow controller with a similar device from the same company capable of regulating the exhaust helium pressure as well. By tuning the PID feedback parameters of the controller, the fluctuations in both temperature and magnetic field could be reduced below the sensitivity of our measurements, as seen in the black curves in Figures \ref{fig:PbTempFFT} and \ref{fig:BzFFT}. 

This stabilizing effect on the magnetic field can also be seen directly in the time domain in Figure \ref{fig:Bz}. Earlier times in the Figure show the stable pressure control mode, and later times are after switching to mass flow rate control to demonstrate the change in fluctuations. A drift of $\gtrsim$ 10 nT/h in the magnetic field does remain over times of $\sim$ 1000 s. This is due to thermoelectric currents generated by differential warming in the warm bore and its thermal shield when nitrogen cooling is removed. The overall temperature drift is seen to be $\sim$ 1 K/h, but different sections of these parts can warm at different rates, especially depending on proximity to liquid cryogen tubes. These currents produce magnetic fields and induce currents in other components, further generating currents and fields. Thus, we can not rule out even the observed sign change in the magnetic drift. Further, the magnitude of the generated fields is not mitigated by magnetic shielding, since there is none between the warm bore and fluxgate magnetometer. If interpreted as a fluctuation in the room, our measured shielding factors (see Table \ref{table:ShieldingFactors}) would imply a field variation in the local field much larger than any observed during running, but this shielding factor does not apply to a field generated in the warm bore itself. 



\begin{figure}[htp]
		\centering
		\includegraphics[width=5.0in]{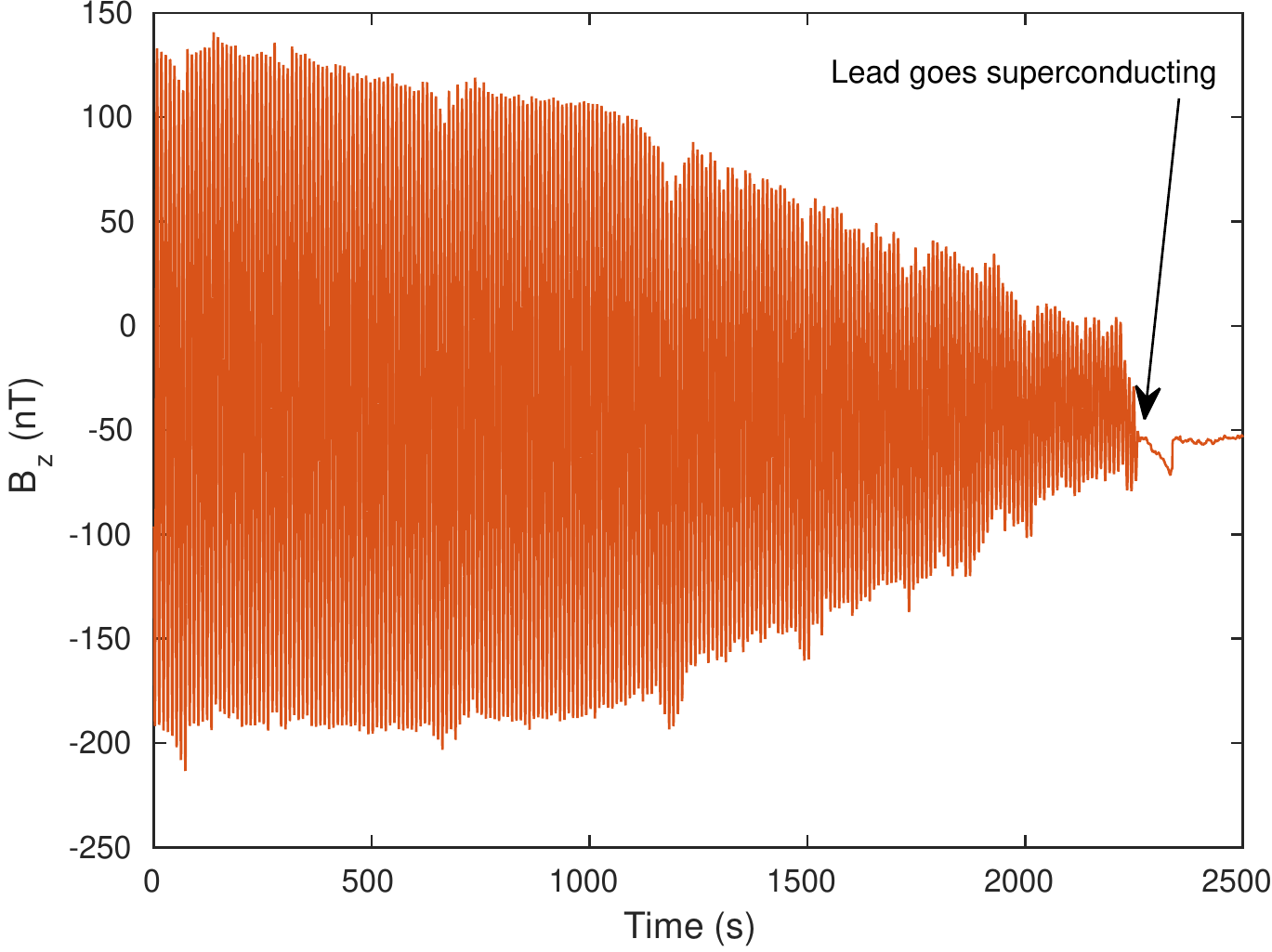}
		\caption{The reduction of noise, as simulated by a 0.1 Hz quasi-static AC magnetic field, as the lead transitions into the superconducting state.}
		\label{fig:LeadgoesSC}
\end{figure}

\newpage
\subsection{Magnetic Shielding}
\label{subsec:Results_shield}

The superconducting shield is effective at mitigating the effects of transient changes in the external magnetic environment. Figure \ref{fig:LeadgoesSC} shows a visualization of the transition of the lead shield to the superconducting state. A 0.1 Hz quasi-static AC signal is generated in the axial compensation coils as a simulation of a noisy environment, and the field is monitored by the probe at the center of the lead shield as it goes from the normal state to superconducting. After the transition a dramatic reduction in the ``noise'' is observed.


To quantitatively characterize the effectiveness of the shield, we apply a quasi-static AC driving field using the compensation coils outside the cryostat. The AC current driving the field is made large enough to be measured with the lead shield in the superconducting state; this is observed to be $B_{\mathrm{unshielded}}=1.44~\mu$T in amplitude at the probe position in the absence of shielding for all frequencies. We measure the field at the center of the shield, then measure the field at the same location after the shield is removed. We can then define a shielding factor, $S$, as the ratio of the two measurements:

\begin{align}
S=\frac{B_\mathrm{unshielded}}{B_\mathrm{shielded}}. \label{eq:SF}
\end{align}

Induced eddy currents and skin depth effects lead to a frequency dependence in the shielding factor, providing enhanced shielding of magnetic fields at higher frequencies. Thus, the quasi-static shielding factor is determined from the asymptotic behavior of the shielding factor at low frequencies. Using Equation \ref{eq:SF}, we measure a total axial shielding factor $S^A=4882\pm174$. The total DC transverse shielding factor is $S^T=79.0\pm3.1$.  Uncertainty in the axial or transverse total shielding factor is determined from the magnitude of the background noise spectrum at the driving frequency.

We are also interested in extracting shielding factors due to the superconducting shield itself, $S_{SC}$, as opposed to the total shielding factor, $S$, which includes other metallic and ferromagnetic components.  In order to do this we measure the magnetic field - $B_N$ - with the lead shield in the normal state above the superconducting transition temperature and - $B_{SC}$ - with the shield in the superconducting state below the transition temperature.  Because of its finite conductivity above the transition temperature, the lead also can shield magnetic fields due to induced eddy currents and skin depth effects. Again, these effects are only important at higher frequencies. Thus, assuming that the various components of the shielding (ferromagnetic and conductor effects) enter multiplicatively, we can extrapolate $B_N/B_{SC}$ to low frequency in order to determine the superconducting shielding factor for the lead. 

However, the analysis is complicated by several factors. First, since we are comparing results at two different temperatures, we want to insure that most of the other shielding is not changed for the two measurements. Thus, we don't want the temperatures to differ by an amount that will cause a significant change in the conductivity, and hence skin depth, of the components beyond the lead. To that end, we performed the $B_N$ measurements with the lead at $T \lesssim 30$ K.  Another complication was the appearance of additional magnetic fields from eddy currents in the conductors near the magnetometer (e.g., the warm bore and LN shield). At higher frequencies this can actually increase the observed magnetic field, especially for axial measurements, due to local oscillating dipoles from the induced currents that are not shielded to the extent of the nearly uniform fields produced by the external coils. However, since this effect should show up for both $B_N$ and $B_{SC}$ measurements it should largely cancel in the ratio of the fields. 


In order to perform this extrapolation, we model the shielding of the normal state lead due to the induced eddy currents. It turns out that this effect dominates over the skin depth effect for the lead due to the large radius ($R=0.38$ m) and small wall thickness ($d=0.8$ mm) - this effect is discussed in Ref.  \cite{fahy1988}. To model this effect we consider the oscillating magnetic field as a source of EMF the drives a series $LR$ circuit, which is the lead conducting cylinder. In the axial case the EMF induces currents circulating azimuthally around the cylinder. For a long cylinder (length $\ell\gg R$) the fields can be calculated analytically as follows. 

Assuming an applied field of $B_0$ varying harmonically as $e^{i\omega t}$, a solenoidal current $I_s$ will be induced which produces a net interior magnetic field of $B_i=B_0+B_s$, where $B_s$ is the field resulting from $I_s$. The EMF is given by Faraday's law

\begin{align}
\mathcal{E} = - \frac{d\Phi}{dt} = -i\omega B_0 (\pi R^2) = I_s Z
\end{align}
where $Z$ is the total impedence $Z=Z_R+Z_L$.

For the axial geometry with $\ell\gg R$ we have $Z_L = i\omega\mu_0\pi R^2/\ell$ and $Z_R = 2\pi R/\sigma d \ell$. This case is a solenoidal current distribution so that the field from the shield is uniform with magnitude $B_s = \mu_0 I_s/\ell$. This gives, using $B_i=B_0+B_s$, the complex ratio 

\begin{align}
\frac{B_i}{B_0} = \frac{1}{1+\frac{iRd}{\delta^2}},
\end{align}
where $\delta$ is the skin depth of the shield given by 

\begin{align}
\delta=\left(\frac{2}{\sigma \mu_0 \omega} \right)^{\nicefrac{1}{2}}, \label{eq:skindepth}
\end{align} 
in agreement with the result of Fahy, Kittel, and Louie \cite{fahy1988}. The sign difference in the denominator is due to our choice of phase. This then gives a reduction of the internal field of 

\begin{align}
\frac{|B_i|}{|B_0|} = \frac{1}{\sqrt{1+\frac{R^2d^2}{\delta^4}}}
\end{align}

For the transverse geometry with $\ell\gg R$ we have $Z_L = i\omega\mu_0\ell/2$ and $Z_R = 2\ell/\sigma d\pi R$. In this case we have a saddle current distribution with a current distribution varying as $\cos\theta$ traveling along the length of the tube, so that again we have a uniform field from $I_s$ given by  $B_s = \mu_0I_{s}/4R$, which gives

\begin{align}
\frac{|B_i|}{|B_0|} = \frac{1}{\sqrt{1+\frac{R^2d^2}{\delta^4}\left (\frac{\pi}{2}\right )^2}}.
\end{align}

In both cases there are finite size effects that modify the geometrical factors, but the frequency dependence should be approximately the same.

In order to extract the quasi-static superconducting shielding factor, a fitted curve of AC measurements is extrapolated to zero frequency. 
As the frequency tends to infinity, the shielding factor for the normal shield should approach that of the superconductor, as the boundary conditions are the same. In order to account for this, we
approximate the normal state shielding factor as the  harmonic mean of the superconducting shielding factor and the $\ell\gg R$ shielding factor calculated above

\begin{align}
\frac{1}{S_{N}}=\frac{1}{S_{SC}}+\left|\frac{B_i}{B_0}\right|, 
\end{align}

Then with $S_N=B_{Pb}/B_N$ and $S_{SC}=B_{Pb}/B_{SC}$ where $B_{Pb}$ is the net field incident on the lead we obtain

\begin{align}
\left(\frac{B_N}{B_{SC}}\right)=1+\frac{S_{SC}}{\sqrt{1+\alpha_{k}^2\omega^2}}\label{eq:axialfitmodel}
\end{align}
where the index $k$ can indicate either axial ($\alpha_{a}$) or transverse ($\alpha_{t}$) shielding. These values are given by 

\begin{align}
\alpha_{a}=\left(\frac{Rd\sigma\mu_{0}}{2}\right)^{2} 
\end{align}

\begin{align}
\alpha_{t}=\left(\frac{\pi Rd\sigma\mu_{0}}{4}\right)^{2}.
\end{align}

\begin{figure}[tbp]
\begin{center}
\includegraphics[width=0.7\textwidth]{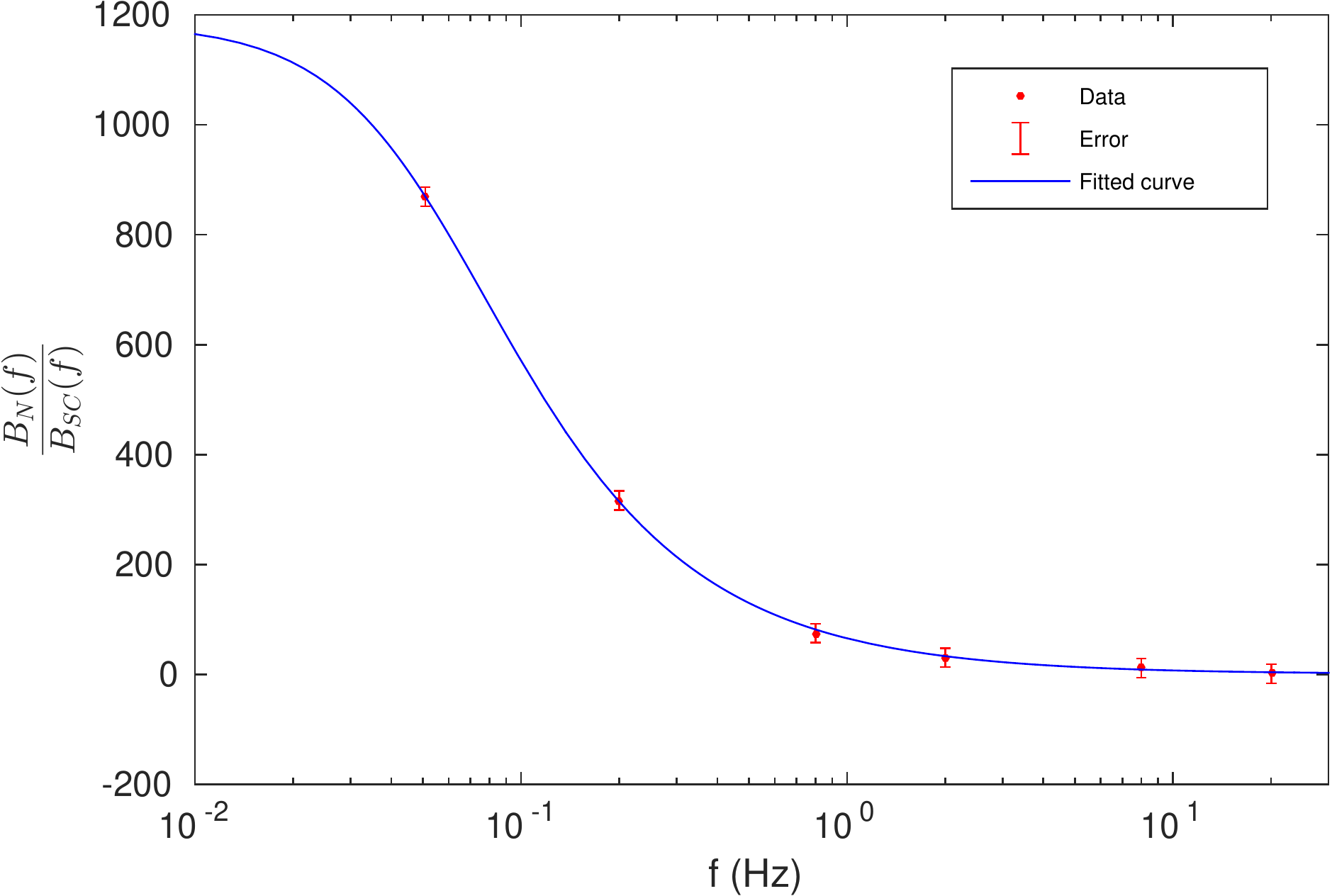}
\end{center}
\caption{The axial superconducting shielding factor is extrapolated from measurements of the lead in superconducting and normal states. The DC value is extrapolated assuming a general skin depth formulation arising from the lead shield in the normal state at $T=33$~K. Error is dominated by the noise with the lead shield in the normal state.}
\label{fig:fittedAxial}
\end{figure}
\begin{figure}[tbp]
\begin{center}
\includegraphics[width=0.7\textwidth]{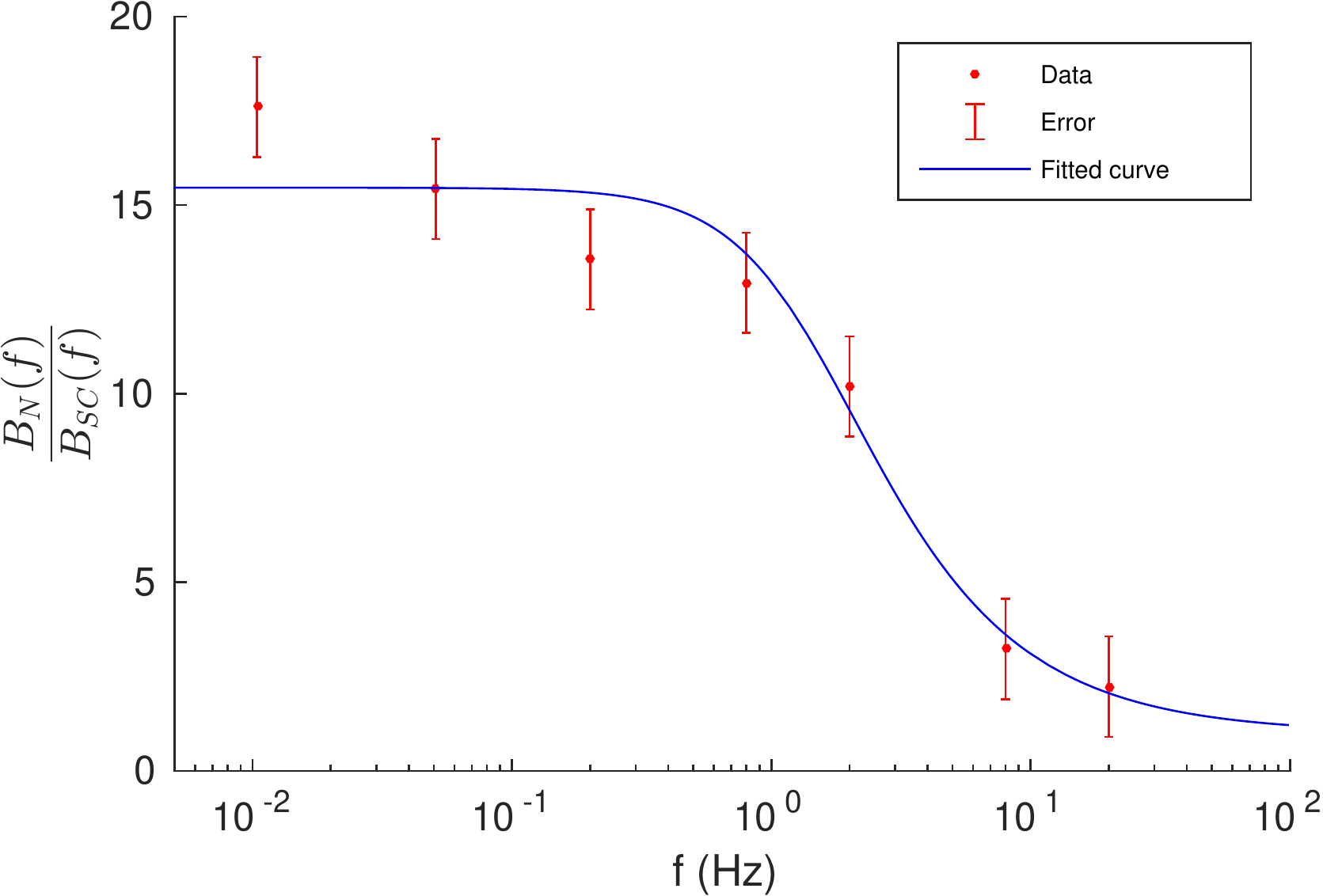}
\end{center}
\caption{The transverse superconducting shielding factor is extrapolated from measurements of the lead in superconducting and normal states at $T \sim 30$~K.  The DC value is extrapolated assuming a skin depth associated with the normal lead shield. Error is dominated by the noise with the lead shield in the normal state. }
\label{fig:fittedTransverse}
\end{figure}

The measurements, along with the best fits using equation \ref{eq:axialfitmodel}, are plotted in Figures \ref{fig:fittedAxial} and 
Figure \ref{fig:fittedTransverse} for the axial and transverse cases respectively. The extracted shielding factors from the best fits are given
in Table \ref{table:ShieldingFactors} along with the total shielding factors discussed earlier. 



\begin{table}[htb]
	\centering
		\begin{tabular}{ccc}
		\hline
	 	 Shielding Factors & Total ($S$) & Superconducting Shield ($S_{SC}$) \\ \hline
		Axial & 4882 $\pm$ 174 & 1183 $\pm$ 131 \\
		Transverse & 79.0 $\pm$ 3.1 & 15.5 $\pm$ 1.2 \\
		\hline	
	\end{tabular}
	\caption{Shielding factors for the entire apparatus and for the superconducting shield alone, as measured at the center of the apparatus with applied AC magnetic fields.}
	\label{table:ShieldingFactors} 

\end{table}


The fit to equation \ref{eq:axialfitmodel} for the transverse case implies a lead conductivity of $\sigma=6\times10^{8}~S/m$, which is consistent with high purity lead around 20 K. The temperature of the lead during the ``normal'' shielding measurements was recorded as 32 K and 24 K on the center and bottom of the cylinder respectively (``Lead Middle'' and ``Lead Bottom'' in Figure \ref{fig:CryoSys}). However, the fit for the axial case implies a much larger conductivity by nearly a factor of 20. This corresponds to temperatures near the superconducting temperature. Since the lead is actively cooled with cold He lines soldered to the surface, it is possible that the circulating induced currents for the axial case sample lead at lower temperatures.

In order to help understand the response of the full system, a model of the apparatus was simulated in COMSOL Multiphysics$^\circledR$ software \cite{COMSOL} at higher frequencies. Due to the limitations in memory the geometry of the model is rudimentary: the warm bore and the warm bore's nitrogen shield are each separately simulated along with the cryostat top flange. Furthermore, smaller features are not simulated; for example the cooling lines on the nitrogen shield and the aluminized mylar used for insulation are larger features that are not included due to memory constraints.

\begin{figure}
\begin{center}
\includegraphics[width=0.7\textwidth]{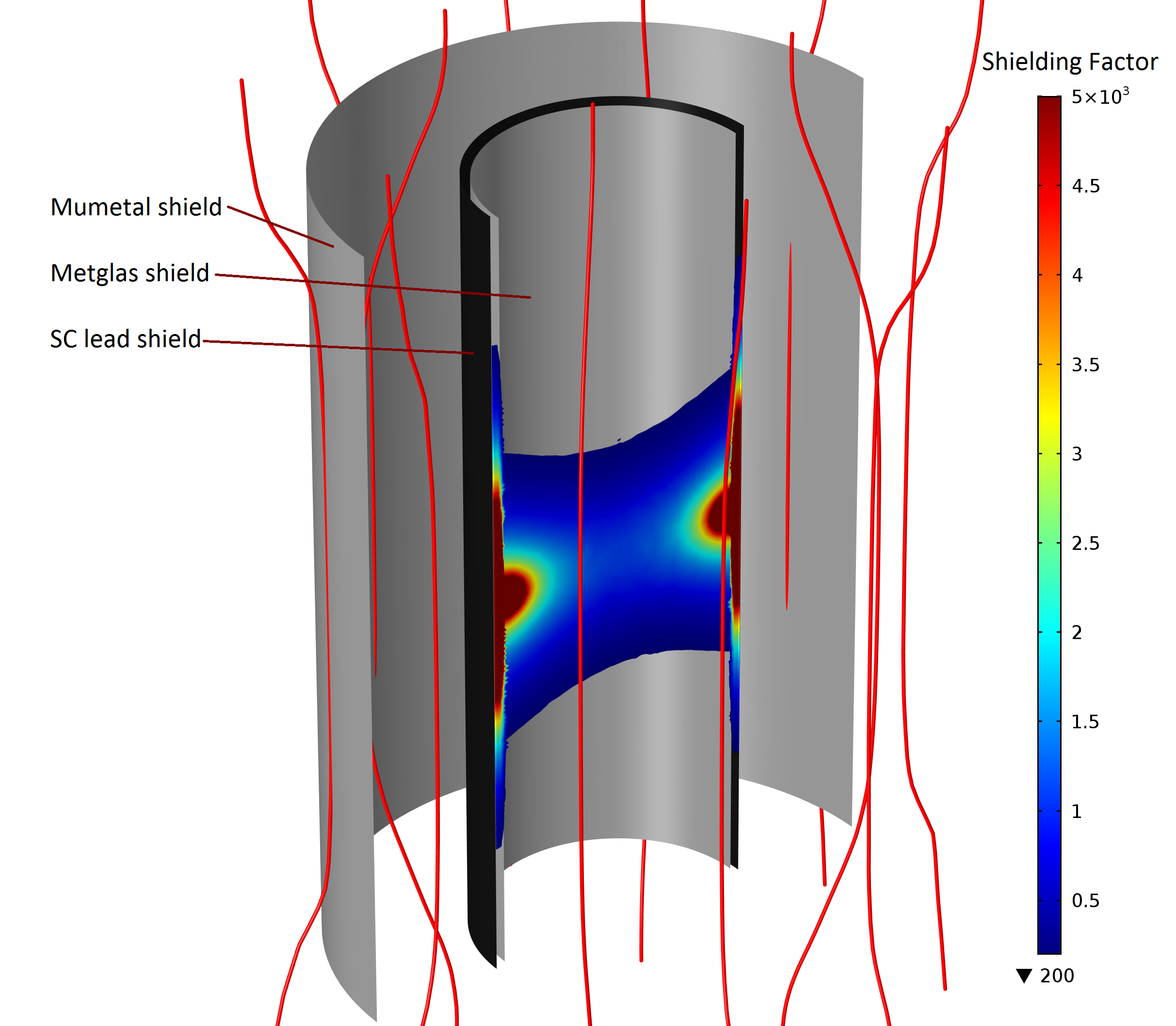}
\end{center}
\caption{Visualization of the DC axial shielding factor in the x-z symmetry plane, as simulated in COMSOL.  Magnetic field lines are shown in red.}
\label{fig:axialshieldCOMSOL}
\end{figure}
\begin{figure}
\begin{center}
\includegraphics[width=0.7\textwidth]{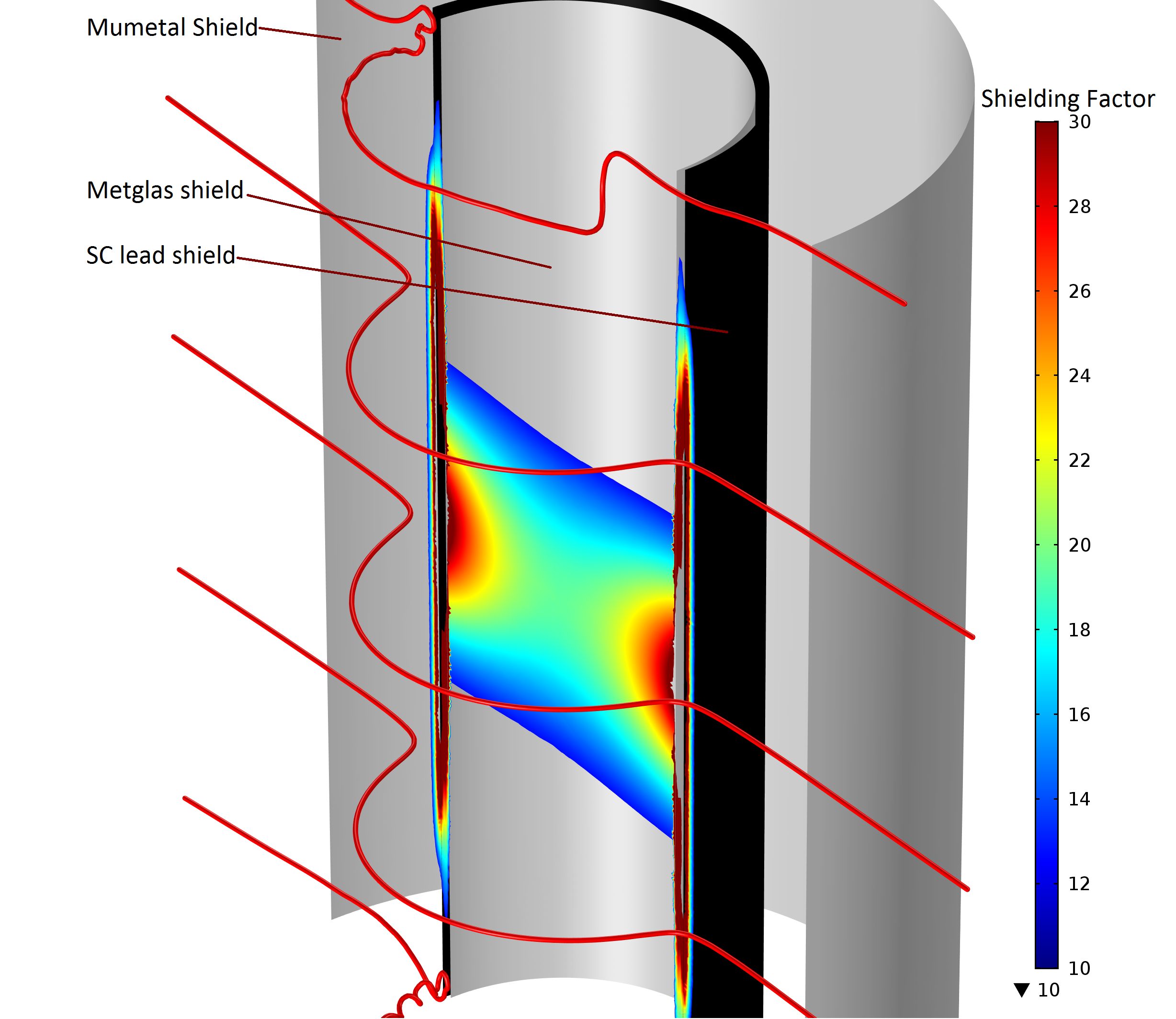}
\end{center}
\caption{Visualization of the DC Transverse shielding factor in the x-z symmetry plane due to the superconducting shield, as simulated in COMSOL. Magnetic field lines are shown in red.}
\label{fig:transshieldCOMSOL}
\end{figure}

The shielding factor is simulated along the x-z symmetry plane of the shielding, seen in cross-section in Figures \ref{fig:axialshieldCOMSOL} and \ref{fig:transshieldCOMSOL}. Also shown in red are the predicted magnetic field lines. Shielding factors are largest nearest the shield itself. Axially, the shielding is weakest at the open ends of the shield and smoothly becomes stronger as the distance from the center decreases. Most relevant for the nEDM experiment is the shielding factor at the center of the $B_{0}$ coil, which is where the measurement cells will be located in the full-scale apparatus. Here, the model predicts shielding factors of $S^{A}_{SC}=1108$ and $S^{T}_{SC}=19.4$ at the center of the $B_{0}$ coil, comparable to the measured values given in Table \ref{table:ShieldingFactors}. 

\subsection{Superconducting endcap results}
\label{subsec:EndcapResults}

The effects of the superconducting endcap were predicted using magnetic simulation 
 to solve for magnetic fields resulting from cylindrically symmetric shields comprised of both ferromagnetic and superconducting materials. The simulation solves for the Green's function with the provided boundary conditions, and proceeds to integrate the Green's function with respect to the requested current density and acquisition points. The calculation includes singular-value decomposition solvers allowing solutions with current densities that circulate on a global geometric scale, an ingredient required for superconducting shields. Small penetrations into the superconducting shield can be accounted for in the simulation by the application of a dipole field at the position of the penetration, with the strength of the dipole calculated according to \cite{jackson}. 

The effects of the endcap are negligible at the center of the fiducial regions, due to the elongated geometry of the magnetic shielding, so it is necessary to map near the region of the endcap. This region is dominated by fringing fields which bow out the open end of the main cylindrical shield, leading to larger vertical ($z$) magnetic field than other regions inside the cylinder. Simulation shows that in the bulk of the lead-enclosed region, a superconducting endcap reduces the $z$-field, leading to more uniform $B_{0}$ field throughout the volume. However, in and above the hole at the center of the endcap, the $z$-field magnitude increases. In effect, the vertical magnetic field is ``pushed out'' of the lead region through the hole in the top.

\begin{figure}[htp]
		\centering
		\includegraphics[width=5.0in]{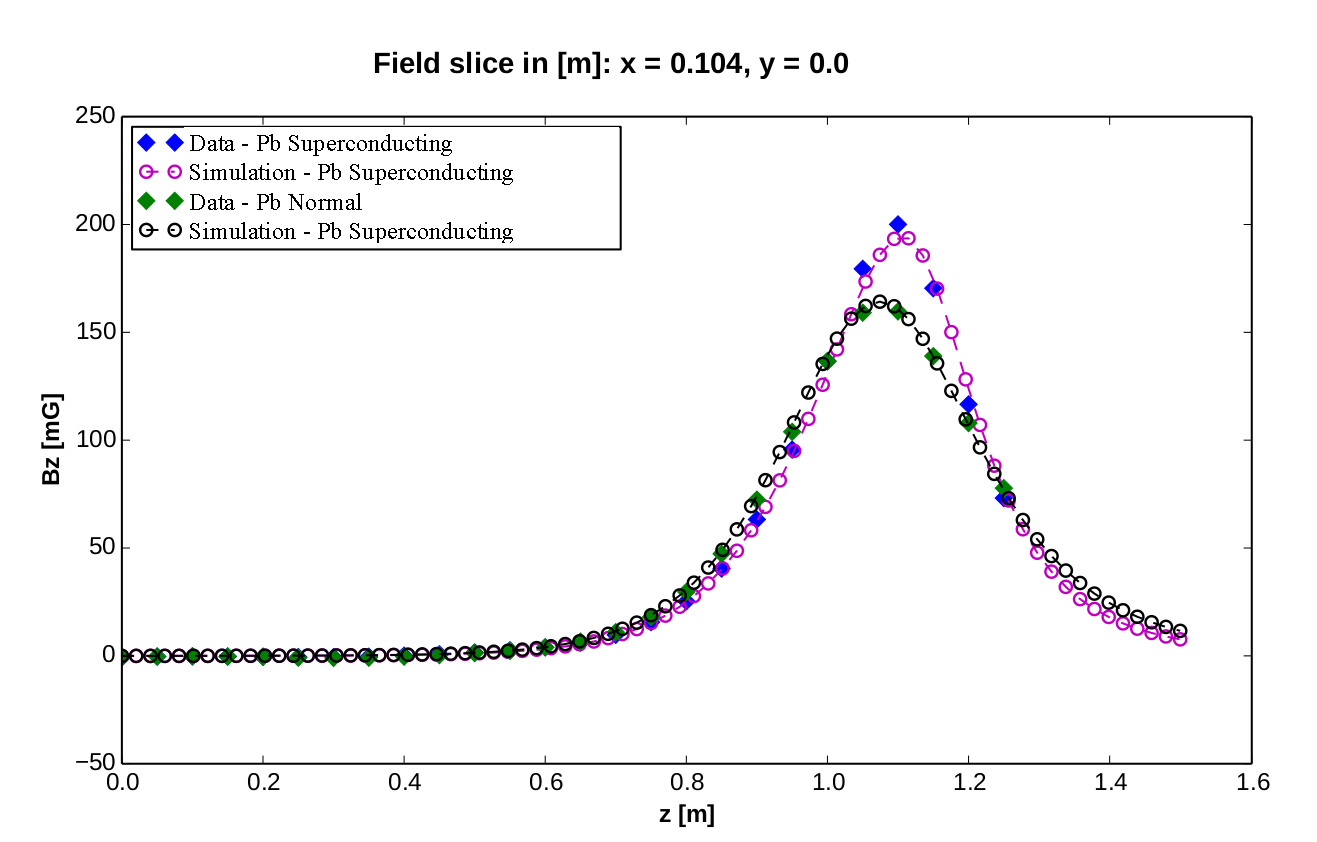}
		\caption{Simulated and measured vertical magnetic field along a vertical slice for superconducting (SC) and normal endcap. The slice is taken off-axis, 0.104 m in the $B_{0}$-direction. The vertical field peaks in the vicinity of the endcap; below the endcap the vertical field is smaller in magnitude when the endcap is SC, indicating greater field uniformity. }
		\label{fig:EndcapFieldSlices}
\end{figure}

Our magnetic probe can be used to map the magnetic field for both the normal and the superconducting state of the lead endcap. The vector magnetic field was sampled along a vertical slice passing from the center of the cell to $\sim$0.5 m above the endcap. The slice was chosen to be 0.104 m away from the axis, in the $B_{0}$ direction, since symmetry prevents any effect at points on the axis. Samples are taken in the presence of a nominal $B_{0}$ field. Measurements are shown in Figure \ref{fig:EndcapFieldSlices} and compared to simulation, which is in excellent agreement. 

\subsection{Magnetic Gradients}
\label{subsec:Results_grad}

Magnetic gradients are measured inside two rectangular fiducial regions in the center of the $B_{0}$ magnet, corresponding to the two measurements cells in the full nEDM experiment, but scaled down to half-size. The regions are 3.8 cm x 5 cm x 20 cm in extent, respectively along the $x$, $y$, and $z$ directions (see Figure \ref{fig:B0Schematic}); they are separated by 5 cm along the $B_{0}$ direction, and are referred to as the ``left'' and ``right'' measurement cells. 
Measurements are taken at night when man-made magnetic activity in the vicinity is reduced. Thermally-driven magnetic fluctuations are mitigated by tuning the liquid helium pressure control and by stopping the flow of liquid nitrogen, as described in Section \ref{sec:ResultsTherm}. The endcap is not in use during maps; it is also too far removed from the cell to have significant effects even if it were superconducting. 

A magnetic ``map'' is defined as a three-dimensional sampling of the fiducial region. The fluxgate magnetometer is moved from point to point along a pre-defined grid, with a travel time of 3-10 s depending on the distance traveled. The magnetic field is measured along 3 orthogonal\footnote{The manufacturer's brochure specifies orthogonality errors of $< 0.1^{\circ}$, which is negligible compared to what is achievable with our alignment procedure.} axes at each point after a settling time of 1 s to reduce vibrational noise. Several measurements of the field are taken over a period of about 1 s and averaged, further mitigating the effects of noise.

Before mapping, the fluxgate is oriented with respect to the $B_{0}$ coil. The vertical (z-direction) center of the $B_{0}$ coil is identified by turning on the coil to the nominal field, centering the probe in x and y by eye, and scanning the probe up and down until a minimum is found. (The $B_{0}$ field is in the x-direction, so the z-field should be smallest when the field is most uniformly in the x-direction.) The probe in the x-direction is then aligned with the $B_{0}$ field by rotating the probe in the x-y plane until the x-field is maximized and the y-field minimized, indicating that the $B_{0}$ field is entirely in the x-direction. By taking the ratio of $B_{y}$ to the $B_{0}$ field, we estimate we can align the probes to about 1\% this way.

Trim coils are then optimized to the current ambient conditions. This is done iteratively in a series of test maps. A coarse map, lasting about 10 min and sampling tens of points, is taken. Trim coils are adjusted by hand to counteract any observed gradients. 

Once trim coils are set, a detailed map of about 5 h is taken with hundreds of points spaced about 1 cm apart. The map is fit using a 3-d quadratic polynomial for each field component. In the 3-d quadratic polynomial model the offsets and coefficients of all possible variations of the three dimensional Cartesian coordinate variables up to second order are fit simultaneously. No constraints are applied. The three probes in the magnetometer are vertically spaced 1.5 cm apart; this distance is accounted for by correcting the position of the measurements before making the fit.

\begin{figure}[htp]
 \centering
   \begin{tabular}{cc}  
   \includegraphics[width={0.55\textwidth}]{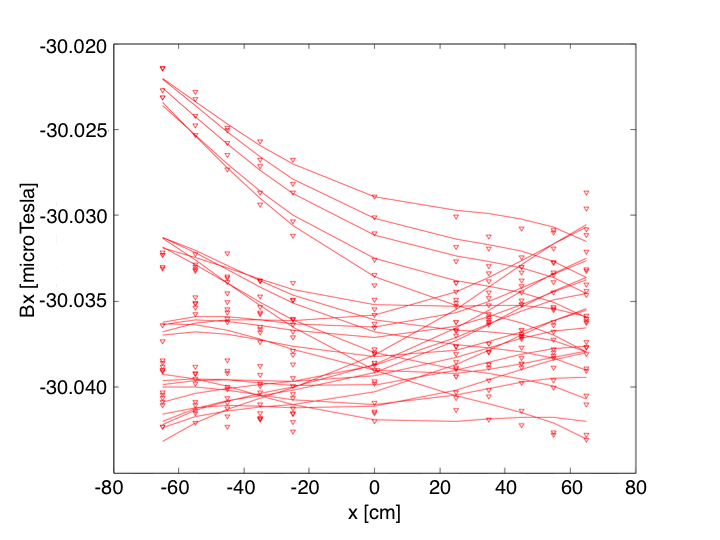}&
   \includegraphics[width={0.5\textwidth}]{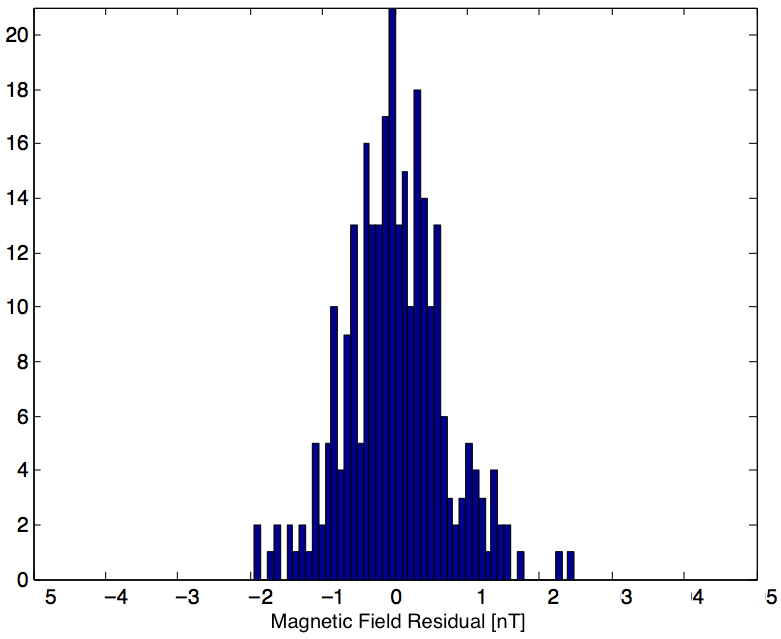}\\  
    \end{tabular}
  \caption{Fits (left) and residuals (right) for an example magnetic map. A quadratic 3-d polynomial fit is used; the figure shows the projection of the fit to $B_{x}$ onto the x-direction. Solid curves indicate series of points taken at fixed values of the y- and z-coordinates. Note that the lead endcap is installed but not in the superconducting state for these measurements.} 
   \label{fig:FitResults}
\end{figure}

Example fits are shown in Figure \ref{fig:FitResults}, along with fit residuals. 
According to Equation \ref{eq:freqshift}, the dominant contribution to the false nEDM is the linear gradient of the magnetic field. Gradients are extracted by taking the partial derivative of the fit model and averaging the position dependent gradients over the fiducial cell volumes. Fit residuals of $\lesssim$1 nT, as compared to the measured field values of $\sim$ 30 $\mu$T, show that statistical uncertainties on the extracted gradients are negligible. The best gradients achieved are shown in Tables \ref{table:BestGradients} and \ref{table:BestOffGradients}, where they have been scaled to the nominal use $B_{0}$ field of 3 $\mu$T and adjusted to the full-scale apparatus.

For this scaling we have assumed that the gradients are dominated by, and thus proportional to, the coil field, $B_{0}$ (noting that the observed gradients are much larger than the ambient gradients). The gradients then scale with the dimensions of the coil and are a factor of 2 smaller for the full-scale system.\footnote{An attempt to verify the scaling by using a smaller, 3 $\mu$T field failed as the measurement was dominated by noise. Larger fields were not possible with our apparatus.} The values listed in Table \ref{table:BestGradients} result in a geometric-phase-induced false dipole moment $d_{f} < 1\times10^{-28}$ e-cm.

\begin{table}[htb]
	\centering
		\begin{tabular}{cccc}
		\hline
		 $<G_{{x}_{i}}>/B_{0} (ppm/cm)$ & $<G_{x}>/B_{0}$ & $<G_{y}>/B_{0}$ & $<G_{z}>/B_{0}$ \\ \hline
		Right Cell & 2.5 & -5.6 & 2.3 \\
		Left Cell & -7.3 & 9.8 & -2.9 \\
		\hline	
	\end{tabular}
	\caption{The best volume-averaged, relative magnetic gradients measured in the prototype apparatus, with $B_{0}$ set to 30 $\mu$T. These terms are most relevant for the false nEDM due to the geometric phase effect. 
	The gradient values have been adjusted to those appropriate for the full-scale experiment. The volume averaging is done separately for the left and right fiducial volumes. 
	Statistical uncertainties are negligible. Note that the superconducting endcap is not in the superconducting state for these measurements.}
	\label{table:BestGradients} 

\end{table}

\begin{table}[htb]
	\centering
		\begin{tabular}{cccc}
		\hline
		 $<G_{x{x}_{i}}>/B_{0} (ppm/cm)$ & $<G_{xy}>/B_{0}$ & $<G_{xz}>/B_{0}$ \\ \hline
		Right Cell & 20.5 & 3.0 \\
		Left Cell & 2.1 & 11.5 \\
		\hline	
	\end{tabular}
	\caption{The best volume-averaged, relative magnetic gradients measured in the prototype apparatus, where $G_{x{x_{i}}} = \partial B_{x_{i}}/\partial x_{i}$ is the gradient of $B_{x}$ in the $x_{i}$ direction. These terms are related to the $^{3}$He transverse polarization relaxation time, $T_{2}$, although the precise dependence is more complicated than a volume average. \cite{McGregor1990}. 
	$B_{0}$ is set to 30 $\mu$T. The gradient values have been adjusted to those appropriate for the full-scale experiment. 
	Statistical uncertainties are negligible. Note that the superconducting endcap is not in the superconducting state for these measurements.}
	\label{table:BestOffGradients} 

\end{table}

\section{Conclusions}
\label{sec:Conclusions}

A prototype magnet system for the nEDM experiment at SNS has been operated cryogenically with minimal thermal disturbance. A superconducting lead shield has been demonstrated to effectively shield external magnetic fields, and field gradients in the prototype are comparable or less than the 3 ppm/cm required by the full experiment. Additionally, a partially superconducting endcap has been investigated experimentally and found to be well modeled by simulation.

Based on these results, we are presently investigating an optimized magnetic design that incorporates superconducting endcaps in a shorter magnetic coil. In this optimized design, the long axis of the measurement cell is oriented perpendicular to the magnet coil axis. This is in contrast with the work presented in this paper, where the long axis is aligned parallel to the coil axis. 

\section*{Acknowledgments}
\label{sec:Acknowledgments}

The authors thank R. Golub, M. Hayden, and S. Lamoreaux for their many useful comments and suggestions. This work was supported by U.S. National Science Foundation grants 1205977 and 1506459.





\newpage
\bibliographystyle{unsrt}
\bibliography{./CryogenicMappingPaper}







\end{document}